\documentclass[a4paper,12pt,reqno]{amsart}
\pdfoutput=1

\usepackage{a4}
\usepackage{amsthm}
\usepackage[T1]{fontenc}
\usepackage[utf8]{inputenc}
\usepackage[british]{babel}
\usepackage{todonotes}
\usepackage[unicode,psdextra]{hyperref}
\usepackage{tikz}
\usetikzlibrary{decorations.pathmorphing,shapes}
\usetikzlibrary{arrows.meta,calc}
\def\centerarc[#1](#2)(#3:#4:#5)
    { \draw[#1] ($(#2)+({#5*cos(#3)},{#5*sin(#3)})$) arc (#3:#4:#5); }

\newtheorem{definition}{Definition}

\newtheorem{theorem}[definition]{Theorem}
\newtheorem{lemma}[definition]{Lemma}

\newtheorem{corollary}[definition]{Corollary}
\newtheorem{proposition}[definition]{Proposition}
\newtheorem{conjecture}[definition]{Conjecture}

\DeclareMathOperator{\Res}{Res}

\makeatletter
\@addtoreset{definition}{section}
\@addtoreset{equation}{section}
\makeatother

\makeatletter
\let\@wraptoccontribs\wraptoccontribs
\makeatother

\allowdisplaybreaks[4]

\begin{document}

\title[Genus one free energy contribution to the quartic Kontsevich 
model]{Genus one free energy contribution to the quartic Kontsevich 
model}

\author[J. Branahl]{Johannes Branahl\textsuperscript{1}}
\author[A. Hock]{Alexander Hock\textsuperscript{2}}
 
\address{\textsuperscript{1}Mathematisches Institut der
  Westfälischen Wilhelms-Universit\"at \hfill \newline
Einsteinstr.\ 62, 48149 M\"unster, Germany \hfill \newline
{\itshape e-mail:} \normalfont
\texttt{j\_bran33@uni-muenster.de}}

\address{\textsuperscript{2}Mathematical Institute, University of Oxford,
  \newline
  Andrew Wiles Building, Woodstock Road, OX2 6GG, Oxford, United Kingdom
  \newline
  {\itshape e-mail:} \normalfont \texttt{alexander.hock@maths.ox.ac.uk}}

\begin{abstract}
  We prove a formula for the genus one free energy $\mathcal{F}^{(1)}$ of the quartic Kontsevich model for arbitrary ramification by working out a boundary creation operator for blobbed topological recursion. We thus investigate the differences in $\mathcal{F}^{(1)}$ compared with its generic representation for ordinary topological recursion. In particular, we clarify the role of the Bergman $\tau$-function in blobbed topological recursion. As a by-product, we show that considering the holomorphic additions contributing to $\omega_{g,1}$ or not gives a distinction between the enumeration of bipartite and non-bipartite quadrangulations of a genus-$g$ surface.  
\end{abstract}


\maketitle
\markboth{\hfill\textsc\shortauthors}{\textsc{{Genus one free energy contribution to the quartic Kontsevich 
model}\hfill}}

\section{Introduction and Main Result}
\subsection{Overview}
This paper derives a formula for the genus one free energy $\mathcal{F}^{(1)}$ of the quartic Kontsevich model (QKM) as a prime example for the explicit construction of $\mathcal{F}^{(1)}$ in blobbed topological recursion (BTR). As a by-product, it gives insights into the interpretation of the holomorphic additions in the context of enumeration of ribbon graphs/quadrangulations of the genus-$g$ surfaces. \\ 
The model got its name by its analogy to the famous Kontsevich model \cite{Kontsevich:1992ti} - both are $N \times N$ Hermitian matrix models with the same covariance, the latter one with a cubic potential, ours with a quartic one. In \cite{Branahl:2020yru} interwoven loop equations (Dyson-Schwinger equations) for three families of correlation functions, among them the expected meromorphic differential forms $\omega_{g,n}$, were solved for low topologies. Their solutions strongly supported the conjecture that the invariants of the QKM obey an extension of Eynard's and Orantin's topological recursion \cite{Eynard:2007kz}: The previous framework enriched by an infinite stack of initial data (\textit{blobs}), successively contributing at each recursion step \cite{Borot:2015hna} - a theory called blobbed topological recursion. These additional quantities are holomorphic at the ramification points of the branched covering $x$ building the recursion's spectral curve, so that one can decompose the  like $\omega_{g,n}(z,u_1,...,u_{n-1})= \mathcal{H}_z \omega_{g,n}(z,u_1,...,u_{n-1})+\mathcal{P}_z \omega_{g,n}(z,u_1,...,u_{n-1})$ into holomorphic and polar part at the ramification points. \\ \\
In \cite{Hock:2021tbl}, the complete genus-0 sector was proved to follow blobbed topological recursion - with a considerable combinatorial effort and, moreover, a remarkable involution identity building on the  symmetry $y(z)=-x(-z)$ of the spectral curve. Its fixed point 0 shall become a key ingredient for $\mathcal{F}^{(1)}$ in BTR. \\ 
Recall that in topological recursion the free energies of stable topology, $\mathcal{F}^{(g>1)}$, are obtained by the application of the primitive of $\omega_{0,1}$ (\textit{dilaton equation}). $\mathcal{F}^{(1)}$ and $\mathcal{F}^{(0)}$  instead are very special: Working out $\mathcal{F}^{(0)}$ was a more or less trivial task (Chap. 5, \cite{Branahl:2020b}), since $\omega_{0,1}$ is not affected by additions due to BTR. The most general form in topological recursion for $\mathcal{F}_{TR}^{(1)}$ for genus zero spectral curves was given in \cite{Eynard:2007kz}
 \begin{align}\label{F1Ey}
\mathcal{F}_{TR}^{(1)}=-\frac{1}{2} \ln[\tau_B(b_1,...,b_{2d})] -\frac{1}{24} \ln \biggl ( \prod_{j=1}^{2d} y'(\beta_j) \biggl )
\end{align}
where $\beta_i$ denote ramification points of the branched covering $x(z)$ ($x'(\beta_i)=0$) and where the Bergman $\tau$-function (introduced in the context of Hurwitz spaces, see \cite{Koko:2003}) is implicitly defined via
 \begin{align}
\label{eq.f1gen}
\frac{\partial }{\partial b_i} \ln[\tau_B(b_1,...,b_{2d})]= \Res_{q \to \beta_i} \frac{B(q,\sigma_i(q))}{dx(q)}
\end{align}
Here we write $x(\beta_i)=:b_i$ and $\sigma_i$ as the local Galois involution nearby $\beta_i$. Solving this differential equation is for many coverings far from trivial - no general solution strategy exists. In the Kontsevich model the $\tau$-function vanishes, in the Hermitian one-matrix model it can be easily solved and in the two-matrix model it was content of a whole paper \cite{Eynard:2002}. This paper shall give an impression for $\mathcal{F}^{(1)}$ in \textit{blobbed} topological recursion: The structure of $\mathcal{F}^{(1)}$ contains then both the well-known term from ordinary TR and a conceptually new term that guarantees the correct generation of the holomorphic part $\mathcal{H}_z \omega_{1,1}(z)$ having poles at $z=0$.

 Moreover, in the case of a single (degenerate) eigenvalue of the external matrix $E$ (the realm of the Hermitian one-matrix model), we show the combinatorial meaning of the blobs contributing to $\omega_{g,1}$ by a perturbative analysis and discuss the role of the Bergman $\tau$-function in our model satisfying blobbed topological recursion.
\subsection{Statement of the Result}
The paper proves the following main theorem:
\begin{theorem}\label{THm1}
The free energy $\mathcal{F}^{(1)}$ reads for the quartic Kontsevich model 
 \begin{align*}
\mathcal{F}^{(1)} = \frac{\mathfrak{R}_{\neq}}{24}-\frac{1}{24} \ln \biggl ( R'(0) \prod_{i=1}^{2d} R'(-\beta_i) \biggl ) 
\end{align*}
where $\beta_i$ are the ramification points of $x(z)=R(z)=-y(-z)$, where $R$ is defined in \eqref{RDef}. Thus, the structure of $\mathcal{F}^{(1)}$ differs by its usual definition \eqref{F1Ey} by the fixed point of the global involution, namely 0, and the $\tau$-function. $\mathfrak{R}_{\neq}$ vanishes for $d=1$ and can be given approximately in terms of the eigenvalues of the external matrix: \footnote{In Chapter \ref{ch:setup} we will review the quartic Kontsevich model in greater detail and specify the background of $R(z)$ and the eigenvalues $e_k$}
\begin{align*}
\mathfrak{R}_{\neq} = -\frac{\lambda}{N} \sum_{k,l, k\neq l}^d \frac{r_kr_l}{(e_k+e_l)^2} + \mathcal{O}(\lambda^2)
\end{align*}
\end{theorem}
In the regime of graph enumeration (one degenerate eigenvalue), we motivate perturbatively and justify with the one- and two-matrix model:
\begin{conjecture}
Let the external matrix in the QKM be a scalar multiple of the identity matrix. Then the holomorphic additions $\mathcal{H} \omega_{g,n}$ have the following interpretation in terms of enumerative geometry:
\begin{itemize}
\item Consider only normalised solutions in the sense of \cite{Borot:2015hna}. Then the pure topological recursion with spectral curve $(x,y)$ of Theorem \ref{THm1} for $\omega_{g,1}^{TR}$ are generating functions for the number of \textbf{bipartite} rooted quadrangulations of a genus-$g$ surface (in the sense of \cite{Carrell:2014}).
\item Let the blobs contribute. Then the complete results for $\omega_{g,1}$ are generating functions for rooted quadrangulations of a genus-$g$ surface, bipartite and non-bipartite ones.
\end{itemize}
\end{conjecture}
For $g=0$, all quadrangulations are bipartite and $\omega_{0,1}^{TR}=\omega_{0,1}$ by definition. The second statement is clear due to the one-to-one correspondence of the quartic Kontsevich model to the Hermitian one-matrix model. \\ \\ 
After briefly introducing the necessary background knowledge regarding the QKM in Ch. \ref{ch:setup}, we start with the combinatorial limit of an \textit{external scalar multiple of the identity matrix} to construct a special case of $\mathcal{F}^{(1)}$ in Ch. \ref{ch:comb}, making use of known results of enumerative geometry. This limit will also give remarkable insights into the characterisation of the generated Feynman graphs/maps generated by pure TR and BTR. In order to derive $\mathcal{F}^{(1)}$ in its full generality (for arbitarily many simple ramification points), we formulate in Ch. \ref{ch:proof} a boundary creation operator for blobbed topological recursion that allows to carry out the proof. We conclude with the remark that an unexpected compensation term can only be given approximately in perturbation theory and that we are grateful for any sort of hints. \\ \\
\centerline{\sc Acknowledgements}

\smallskip

We thank Jakob Lindner for his numerical support of our perturbative approximations and Raimar Wulkenhaar for helpful comments. JB is supported\footnote{``Funded by
  the Deutsche Forschungsgemeinschaft (DFG, German Research
  Foundation) -- Project-ID 427320536 -- SFB 1442, as well as under
  Germany's Excellence Strategy EXC 2044 390685587, Mathematics
  M\"unster: Dynamics -- Geometry -- Structure."} by the Cluster of
Excellence \emph{Mathematics M\"unster}. AH is supported through
the Walter-Benjamin fellowship\footnote{``Funded by
  the Deutsche Forschungsgemeinschaft (DFG, German Research
  Foundation) -- Project-ID 465029630}.

{\footnotesize\tableofcontents}

\section{The Quartic Kontsevich Model}
\label{ch:setup}
The original aim of the quartic Kontsevich model was to construct an exactly solvable quantum field theory on non-commutative spaces, where the property of exact solvability might be explained by the appearance of topological recursion. We will not go into the details of the background (and refer the interested reader to a new overview of the accomplishments in the last two decades, to find in \cite{Branahl:2021}) and only take concepts into account that will be necessary to understand this paper. Consider an integral over self-adjoint $N\times N$-matrices living in $H_N$
\begin{align}
\mathcal{Z}=\frac{\int_{H_N}d \Phi   \exp \big ( - N \mathrm{Tr}(E \Phi^2 +\frac{\lambda}{4}\Phi^4  ) \big ) }{\int_{H_N}d \Phi   \exp \big ( - N \mathrm{Tr}(E \Phi^2)\big )}
\label{partfunc}
\end{align}
with a scalar $\lambda$ and a Hermitian $N\times N$-matrix $E$ with positive eigenvalues. From this \textit{partition function} of the quartic Kontsevich model (the original Kontsevich model carries a $\frac{\lambda}{3}\Phi^3 $) we investigate the following moments (decomposing into cumulants)
\begin{align}
\langle  \Phi_{k_1l_1}...  \Phi_{k_nl_n}\rangle 
&=\frac{1}{N}\frac{\int_{H_N}d \Phi  \Phi_{k_1l_1}...  \Phi_{k_nl_n} \exp \big ( - N \mathrm{Tr}(E \Phi^2 +\frac{\lambda}{4}\Phi^4  ) \big ) }{\int_{H_N}d \Phi   \exp \big ( - N \mathrm{Tr}(E \Phi^2+\frac{\lambda}{4}\Phi^4)\big )},
\label{cumulants}
\end{align}
where $\Phi_{kl}$ denotes the entry of the matrix $\Phi$.
We give a brief sketch about the main ingredients we needed to recognize the structure of \textit{blobbed topological recursion (BTR)} in our model (more details to find in \cite{Branahl:2020yru}). Cumulants of the particular form
\begin{align}
N^{n_1+...+n_b} 
\big\langle ( \Phi_{k_1^1k_2^1} 
 \Phi_{k_2^1k_3^1} ... 
 \Phi_{k_{n_1}^1k_1^1}) ... 
( \Phi_{k_1^bk_2^b}  \Phi_{k_2^bk_3^b} ... 
 \Phi_{k_{n_b}^bk_1^b}) \big\rangle_c 
=: N^{2-b} \, G_{|k_1^1... k_{n_1}^1|...
|k_1^b... k_{n_b}^b|} \;,
\label{GDef}
\end{align}
are defined to be \textit{correlation functions} (precisely: $(n_1{+}...{+}n_b)$-point functions) having a formal $1/N$-expansion
$G_{|k_1^1... k_{n_1}^1|...  |k_1^b... k_{n_b}^b|} =:
\sum_{g=0}^\infty N^{-2g} G^{(g)}_{|k_1^1... k_{n_1}^1|...
  |k_1^b... k_{n_b}^b|} $.   The $N$ eigenvalues later decay into $d$ distinct eigenvalues with multiplicity $r_1$ to $r_d$ (technically spoken, this is only allowed after the complexification of the correlation functions, but for the purposes of the paper, we forget about this subtlety).

The starting point of the recursion that governs the quartic Kontsevich model is the introduction of a function $\Omega_{q} := \frac{1}{N}\sum_{k=1}^N G_{|qk|}+\frac{1}{N^2}G_{|q|q|}$, again having a genus expansion, made of the simplest correlation functions. The key step towards (B)TR was the definition of a \textit{boundary creation operator} that will be the main tool throughout the paper:
\begin{definition}
Define the boundary creation operator as 
\begin{align}
\hat T_b := -\frac{N}{r_b} \frac{\partial}{\partial e_b}\; .
\label{eq:creator}
\end{align}
 The free energies are the primitives of $\Omega_{q}^{(g)}$ under $\hat T_q$, i.e. $\hat T_q \mathcal{F}^{(g)}:= \Omega_{q}^{(g)}$, or equivalently $\sum_{g=0}^\infty N^{2-2g}\mathcal{ F}^{(g)}=\log \mathcal{Z}$.
\end{definition}
We will later focus on the explicit computation $\hat T_q \mathcal{F}^{(1)}= \Omega_{q}^{(1)}$. The boundary creation operator gives rise to $\Omega_{q_1,...,q_m}$ (symmetric in their indices) in the same manner:
\begin{align}\label{Om}
\Omega_{q_1,...,q_m} &:= 
\hat T_{q_2}...\hat T_{q_m}\Omega_{q_1} 
 +\frac{\delta_{m,2}}{(e_{q_1}-e_{q_2})^2}\;,\qquad
m\geq 2\;.
\end{align}
Applying the boundary creation operator on the correlation functions $G$ defined in \eqref{GDef}, the \textit{generalised correlation functions} are defined by
\begin{align}\label{T}
T_{q_1,q_2,...,q_m\|k_1^1...k_{n_1}^1|k_1^2...k_{n_2}^2|...|k_1^b...k_{n_b}^b|}
:=\hat T_{q_1}...\hat T_{q_m}G_{|k_1^1...k_{n_1}^1|k_1^2...k_{n_2}^2|...|k_1^b...k_{n_b}^b|},
\end{align}
one can also obtain the $\Omega_{q_1,...,q_m}$ via a triangle structure of interwoven Dyson-Schwinger equations (compare with Ch. 5 in \cite{Branahl:2020yru}). All these objects will be later analytically continued as complex-valued meromorphic functions. Evaluating these analytically continued functions at the \textit{energies} (the eigenvalues $e_k$) will provide the $\Omega$'s and $T$'s of \eqref{Om} and \eqref{T}.   \\ \\ 
Starting point towards blobbed topological recursion was the solution of a nonlinear integral equation \cite{Grosse:2009pa} determining the simplest correlation function $G_{|ab|}^{(0)}=G^{(0)}(e_a,e_b)$ which extends to a holormorphic function $G^{(0)}(\zeta, \eta)$. A solution became possible via a variable transform $\zeta \mapsto z=R^{-1}(\zeta)$ \cite{Grosse:2019jnv} \cite{Schurmann:2019mzu} with a biholomorphic mapping $R(z)$ from the right half plane to the complex plane that deforms the eigenvalues $e_k$ implicitly to $\varepsilon_k$ as follows:
\begin{align}\label{RDef}
R(z)&=z-\frac{\lambda}{N} \sum_{k=1}^d \frac{\varrho_k}{\varepsilon_k+z}\;,\qquad
R(\varepsilon_k)=e_k\;,\quad
\varrho_k R'(\varepsilon_k)=r_k \;.
\end{align}
It is convenient to introduce the transformed correlation functions $\mathcal{G}^{(0)}(z,w)=G^{(0)}(R(z),R(w))$. In the same manner, we can complexify and transform the cumulants of an arbitrary topology:
\begin{align*}
\mathcal{G}^{(g)}(z_1^1,...,z^1_{n_1}&|\dots|
z_1^b,...,z^b_{n_b})\\
&:=
G^{(g)}(R(z_1^1),...,R(z^1_{n_1})|\dots|
R(z_1^b),...,R(z^b_{n_b})) \;.
\end{align*}
The variable transform $R$ is the key ingredient of our recursion formula that will create meromorphic differentials $\omega_{g,n}$ defined by
\begin{align*}
\omega_{g,n}(z_1,...,z_n):=\lambda^{2-2g-n}\Omega^{(g)}_{n}(z_1,...,z_n)
dR(z_1)\cdots dR(z_n)
\end{align*}
for negative Euler characteristic, after a similar complexification of $\Omega^{(g)}_{q_1,...,q_n}=:\Omega^{(g)}_{n}(\varepsilon_{q_1},...,\varepsilon_{q_m})$ . The $\omega_{g,n}$ will be generated via the spectral curve
$(x:\hat{\mathbb{C}}\to \hat{\mathbb{C}},
\omega_{0,1}=ydx,\omega_{0,2})$  where $x$ and $y$ are constructed from $R$ defined in \eqref{RDef} with a global symmetry
\begin{align*} 
  x(z)=R(z)\;,\qquad
  y(z)=-R(-z)\;,\qquad
  \omega_{0,2}(u,z)=\frac{du\,dz}{(u-z)^2}+\frac{du\,dz}{(u+z)^2}\;.
\end{align*}
We underline the appearance of some additional initial data
in $\omega_{0,2}$, namely $\frac{du\,dz}{(u+z)^2}=-B(u,-z)$
(Bergman kernel with one changed sign) that will play the role of the first \textit{blob}. The solution of the Dyson-Schwinger equations for some easy topologies $(g,n)$ suggested that not the complete $\omega_{g,n}$ can be created via usual topological recursion, but only the part having poles at the ramification points $\beta_i$, defined via $dR(\beta_i)=0$ (we call it polar part). We recall the well-known recursion formula \cite{Eynard:2007kz}: 
\begin{align}\label{TR}
& \omega_{g,n+1}^{TR}(I,z)
  \\\nonumber
  & :=\sum_{\beta_i}
  \Res\displaylimits_{q\to \beta_i}
  K_i(z,q)\bigg(
  \omega^{TR}_{g-1,n+2}(I, q,\sigma_i(q))
  +\hspace*{-1cm} \sum_{\substack{g_1+g_2=g\\ I_1\uplus I_2=I\\
            (g_1,I_1)\neq (0,\emptyset)\neq (g_2,I_2)}}
  \hspace*{-1.1cm} \omega^{TR}_{g_1,|I_1|+1}(I_1,q)
  \omega^{TR}_{g_2,|I_2|+1}(I_2,\sigma_i(q))\!\bigg)\;.
  \nonumber
\end{align}
with the \textit{local Galois involution} $\sigma_i\neq \mathrm{id}$
  defined via $x(q)=x(\sigma_i(q))$ near $\beta_i$ and  the \textit{recursion kernel} $K_i(z,q)
  =\frac{\frac{1}{2}\int^{q}_{\sigma_i(q)}
    B(z,q')dq'}{\omega_{0,1}(q)-\omega_{0,1}(\sigma_i(q))}$  constructed
  from the initial data with $\omega_{0,2}^{TR}(u,z)=B(u,z)$. 

 Our solutions contain holomorphic add-on's as well, having poles elsewhere. Such an appearance of additional holomorphic parts perfectly fits into the philosophy of \textit{blobbed topological recursion} in which the (so far: planar) holomorphic part $\mathcal{H}_z \omega_{g,n}(z,u_1,...,u_{n-1})$ from $\omega_{g,n}(z,u_1,...,u_{n-1})= \mathcal{H}_z \omega_{g,n}(z,u_1,...,u_{n-1})+\mathcal{P}_z \omega_{g,n}(z,u_1,...,u_{n-1})$ can be generated from a simple recursion formula, remarkably similar to usual topological recursion. 
 The polar part is given by \cite{Borot:2013lpa}:
 \begin{align}\label{TRP}
 	& \mathcal{P}_z \omega_{g,n+1}(I,z)
 	\\\nonumber
 	& =\sum_{\beta_i}
 	\Res\displaylimits_{q\to \beta_i}
 	K_i(z,q)\bigg(
 	\omega_{g-1,n+2}(I, q,\sigma_i(q))
 	+\hspace*{-1cm} \sum_{\substack{g_1+g_2=g\\ I_1\uplus I_2=I\\
 			(g_1,I_1)\neq (0,\emptyset)\neq (g_2,I_2)}}
 	\hspace*{-1.1cm} \omega_{g_1,|I_1|+1}(I_1,q)
 	\omega_{g_2,|I_2|+1}(I_2,\sigma_i(q))\!\bigg)\;,
 	\nonumber
 \end{align}
where we emphasise the omission of $(TR)$ on the rhs of the formula for $\omega$.
 The proof in \cite{Hock:2021tbl} for our $\omega_{0,n+1}(I,z)$ contains that the linear and quadratic loop equations (see \cite{Borot:2013lpa}) are fulfilled, too. Our case of interest, $\omega_{1,1}(z)$, has a holomorphic part with poles at the fixed point of the global symmetry between $x$ and $y$, being 0. Therefore it is natural to assume that $\mathcal{F}^{(1)}$ has to consist of more terms we are going to identify. \\ \\ 
 Finally some words dedicated to approximative solutions: All correlation functions possess a perturbative expansion in the \textit{coupling} $\lambda$ into ribbon graphs/fat graphs, whereas the free energies $\mathcal{F}^{(g)}$ give rise to closed graphs (vacuum graphs). We will not give the complete background of perturbation theory and refer instead to the detailed explanation and illustration in \cite{Branahl:2020b}, where a first perturbative analysis of the QKM was performed. However, we will need the ribbon graph expansion of $\mathcal{F}^{(1)}$ for later purposes, which is outsourced to App. \ref{appA} so that we recapitulate the needed tools:

One can give a meaning to the graph expansion of $\mathcal{F}^{(g)}$ by expanding the exponential in 
  $\log \mathcal{Z}$ in $\lambda$. Let $\mathfrak{G}^{g,v}_\emptyset$  be the set of connected
  vacuum ribbon graphs of genus $g$ made of $v$ four-valent vertices
  and no one-valent vertices (see \cite{Branahl:2020b} for more details). Then we can expand $ \log \mathcal{ Z}$ in the following manner (similar to the discovery of \cite{tHooft:1973alw,Brezin:1977sv}):
\begin{align}     
         \sum_{g=0}^\infty\sum_{v=0}^\infty 
          \sum_{\Gamma_0\in\mathfrak{G}^{g,v}_\emptyset}
          \frac{N^{2-g}\varpi(\Gamma_0)}{|\mathrm{Aut}(\Gamma_0)|}\;,
\end{align} 
where the weights of the Feynman graph $\varpi(\Gamma_0)$ are given as follows:
\begin{itemize}\itemsep -1pt
\item every 4-valent ribbon-vertex carries  a factor $-\lambda$ 
\item every ribbon with strands labelled by $p,q$ carries a factor $\frac{1}{e_p+e_q}$ (\textit{propagator})
\item multiply all factors and apply the summation operator
  $\frac{1}{N^{v+1}}\sum_{l_1,..,l_{v+1}=1}^N r_{l_1}...r_{l_{v+1}}$ over the $v+1$ loops
  (closed strands) labelled by $l_1,...,l_{v+1}$. 
\end{itemize}
In terms of the free energy, we can perturbatively
establish
\begin{align}
  \mathcal{F}^{(g)}= -\frac{\delta_{g,0}}{2N^2} \sum_{k,l=1}^N \log(E_k+E_l)
  +\sum_{v=1}^\infty  \sum_{\Gamma_0\in\mathfrak{G}^{g,v}_\emptyset}\frac{\varpi(\Gamma_0)}{
    |\mathrm{Aut}(\Gamma_0)|} \;.
  \label{freeenergy}
\end{align}
To justify our exact results once more in a perturbative way, we will investigate $\mathfrak{G}^{g,v}_\emptyset$ for $v=1,2$ in App. \ref{appA}.

\section{A Combinatorial Interpretation of BTR}
\label{ch:comb}
Before we give the general proof of our statement, set for this section the number of distinct eigenvalues of the external matrix for simplification to one, that is $d=1$, where we enter the regime of enumerative geometry and combinatorics.

\subsection{$\mathcal{F}^{(1)}$ in the combinatorial limit}
Comparing with the partition function
\begin{align*}
Z_{1MM} = \frac{\int_{H_N} dM \exp \big [ -N \mathrm{Tr}( \frac{M^2}{2} -V(M)) \big ]}{\int_{H_N} dM \exp \big [ -N \mathrm{Tr}( \frac{M^2}{2}) \big ]}
\end{align*}
we obviously reach the Hermitian one-matrix model (1MM) \cite{Brezin:1977sv,Eynard:2016yaa} limit of the quartic Kontsevich model if $V(M)=\sum_{k=3}^dt_k\frac{M^k}{k}$ only contains a quartic term $t_4\frac{M^4}{4}$, giving rise to many helpful cross checks. The special case of interest in the  1MM consists of setting all non-tetravalent vertices to zero, $t_k \mapsto t_k \delta_{k,4}$. In that case, the $\mathcal{F}^{(g)}_{1MM}$ are equivalent to those of the QKM ($t_4=-\lambda$), although it is completely invisible in the exact results due to the different spectral curves and the additional \textit{blobs} in the QKM. However, at the level of graphs/maps, the Feynman rules of our model give exactly rise to the closed graphs of the 1MM with $t_k \mapsto t_k \delta_{k,4}$. The equivalence also holds for $\mathcal{T}_2^{(g)}$ (where the boundary length reads 2 and all other faces are quadrangulations, see its definition in \cite{Eynard:2016yaa}, p. 118) in the 1MM and $\omega_{g,1}$ in the QKM because of the same results produced by the boundary creation/loop insertion operator:
\begin{itemize}
\item in the QKM the creation operator simply becomes $-\frac{\partial}{\partial e}$ and acts on the power series in $\frac{\lambda^{k}}{e^{2k'}}$
\item in the 1MM the creation operator becomes for our purposes $-4 \lambda \frac{\partial}{\partial \lambda}$ since $-4 \frac{\partial}{\partial \lambda} \mathcal{F}^{(g)}_{1MM} = \mathcal{T}_4^{(g)} $ and $[\lambda^l] \mathcal{T}_4^{(g)}=[\lambda^{l+1}] \mathcal{T}_2^{(g)}$ 
\end{itemize}
Deriving $\mathcal{F}^{(g)}$ with respect to the coupling $\lambda$ or to the trivial decoration $e$ of the graphs gives the same results.

 Only for $d=1$, the system of equations determining $R(z)$ can be explicitly inverted (they can be extracted from Chapter 5.3 of \cite{Branahl:2020b}, where some combinatorial illustrations were given for the planar sector). Thus, we had the chance to guess the additional contributions to $\mathcal{F}^{(1)}$ in the QKM due to blobbed topological recursion by comparing their power series. This educated guess became the basis for the general proof. We preliminarily note:
\begin{proposition}
\label{f1d1}
The genus one free energy of the quartic Kontsevich model reads in the special case of a twofold branched cover:
\begin{align}
\label{f1numbers}
\mathcal{F}^{(1)}_{QKM} \underset{d=1}{=}  -\frac{1}{24} \ln   [ R'(0)  R'(-\beta_1) R'(-\beta_2) ]=    \frac{1}{12} \sum_{n=1}^{\infty} \frac{3^n}{n} \biggl ( 2^{2n-1}-\frac{(2n-1)!}{n!(n-1)!} \biggl ) \frac{(-\lambda)^n}{(2e)^{2n}} \nonumber \\
= -\frac{\lambda}{4(2e)^2}+\frac{15}{8(2e)^4}\lambda^2-\frac{33}{2(2e)^6}\lambda^2+\frac{2511}{16(2e)^8}\lambda^4-\frac{15633}{10(2e)^{10}}\lambda^5+...
\end{align}
\end{proposition}
This is as announced the result of \cite[p. 117]{Eynard:2016yaa} for counting closed genus-1 maps with only tetravalent vertices (or only quadrangles as faces, the dual picture) in the Hermitian 1MM. In this model, the exact result reads 
\begin{align*}
\mathcal{F}^{(1)}_{1MM}= -\frac{1}{2} \ln(\tau_B)-\frac{1}{24} \ln \big( y'(1)y'(-1) \big )=\frac{1}{12}\ln \big ( \frac{1+\tilde \lambda}{2\tilde \lambda} \big ) 
\end{align*}
 with $\tilde \lambda = \sqrt{1+12\lambda}$. \textit{Note that this result includes the Bergman $\tau$-function in contrast to the QKM}.  The spectral curve $(x,y)$ is given in the 1MM as 
\begin{align*}
x(z)= \gamma \bigg (z+\frac{1}{z} \bigg ) \qquad y(z)=\frac{1}{\gamma z}+\frac{\lambda \gamma^3}{z^3} \qquad \gamma^2 =  \frac{2}{1+\sqrt{1+12\lambda}}
\end{align*}
   The fixed point of the global symmetry of $x$ and $y$ compensates the differences in the spectral curve (no symmetry in the 1MM) and the $\tau$-function. However, $x(z)$ is very close to the QKM using the following variable transform:
\begin{align*}
&x(z) =R(z)= z- \frac{\lambda}{N} \frac{\varrho}{z+\varepsilon} \quad \Rightarrow  \quad R(t(z)) =-\varepsilon+ \tilde \gamma \bigg (t+\frac{1}{t} \bigg )  \qquad y(z)=-x(-z)\\ 
&\tilde \gamma = \sqrt{-\frac{\lambda \varrho}{N}} \qquad t(z) = \frac{\varepsilon+z}{\tilde \gamma }
\end{align*}
 We recall that the series from above creates rational, not natural numbers $\frac{3^n}{n}  ( 2^{2n-1}-\frac{(2n-1)!}{n!(n-1)!} )$.
These rational coefficients exists due to non-trivial automorphism groups of closed maps (we prefer their duals, ribbon graphs or \textit{fat Feynman graphs} due to our proximity to quantum field theory). 

\begin{figure}[h!]
  \centering
    \includegraphics[width= 0.99\textwidth]{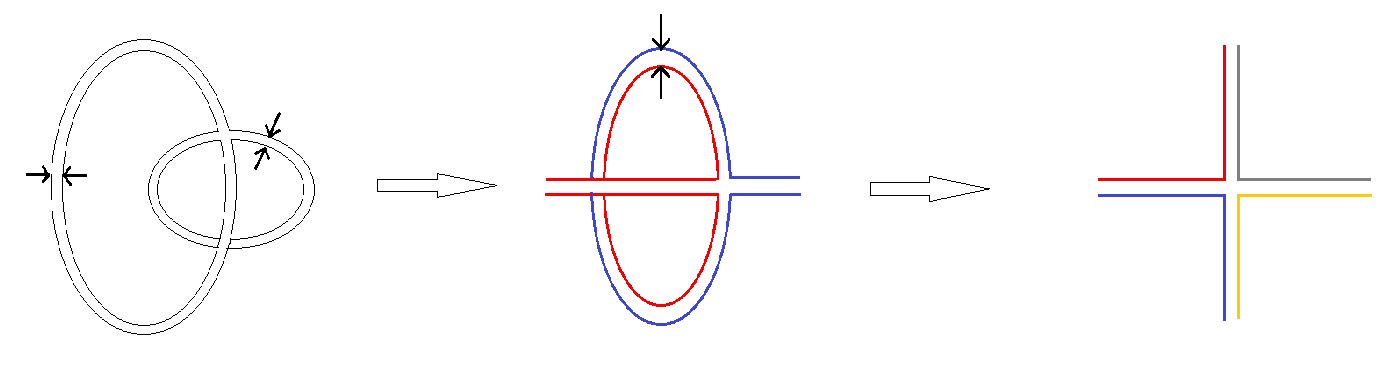} 
    \caption{The $\mathcal{O}(\lambda)$ contribution to $\mathcal{F}^{(1)}$ is cutted by the action of the boundary creation operator causing two faces in the $ \mathcal{G}^{(0)}(\varepsilon|\varepsilon)$ contribution (blue and red). A further cut gives the simplest graph to the four-point function, where $\hat T$ is only allowed to cut the closed loop (more details to be found here: \cite{Branahl:2020b}).
\label{graph1}}
\end{figure}

This topic is a good starting point for the perturbative viewpoint of $\hat T$: The main tool for the general proof of $\mathcal{F}^{(1)}$ is the application of the \textit{boundary creation operator} $\hat T$. It has a very intuitive meaning in the context of ribbon graphs as the perturbative expansions of the correlation functions. Fig. \ref{graph1} shows its action on the $\mathcal{O}(\lambda)$ contribution to $\mathcal{F}^{(1)}$. 
Due to the high symmetry of the graph, their are four ways to cut the ribbon graph leading always to the same graph in $\Omega_{1,1}(\varepsilon)=\mathcal{G}^{(1)}(\varepsilon,\varepsilon) + \mathcal{G}^{(0)}(\varepsilon|\varepsilon)$, more precisely to the 1+1-point function $\mathcal{G}^{(0)}(\varepsilon|\varepsilon)$ as indicated by the colour.  

For $d=1$, the action of the boundary creation operator on $\mathcal{F}^{(1)}_{QKM}$ is easy. Carrying out the derivative  $-\frac{\partial}{\partial e}$ directly gives important natural numbers 
\begin{align*}
\biggl \{ \frac{3^n}{6} \biggl ( 2^{2n}-\frac{(2n)!}{n!n!} \biggl ) \biggl \}_{n\in \mathbb{N}} = \{1,15,198,2511,31266,...\} 
 \end{align*}
counting rooted quadrangulations of the torus (see the accordance with \cite[p. 101]{Eynard:2016yaa}). We recall from \cite{Branahl:2020yru}
\begin{proposition}
\label{om11}
$ \lambda \Omega_{1,1}(z)R'(z)dz =\omega_{1,1}(z) $ reads in the QKM:
\begin{align}
  \lambda \Omega_{1,1}(z)R'(z)
&=
-\frac{\lambda}{8 (R'(0))^2 z^3}
+\frac{\lambda R''(0)}{16(R'(0))^3z^2}
-\sum_{i=1}^{2d} 
\frac{\lambda}{8 \beta_i^2 R''(\beta_i) R'(-\beta_i)  (z-\beta_i)^2}
\nonumber
\\
& + \lambda \sum_{i=1}^{2d} 
\Big\{
-\frac{1}{8 R'(-\beta_i) R''(\beta_i) (z-\beta_i)^4} 
+ \frac{R'''(\beta_i)}{24 (R''(\beta_i))^2 R'(-\beta_i) (z-\beta_i)^3}
\nonumber
\\
&+\frac{R'''(-\beta_i)}{48 R''(\beta_i) (R'(-\beta_i))^2 (z-\beta_i)^2}
+\frac{R'''(\beta_i) R''(-\beta_i)}{48 (R''(\beta_i))^2 (R'(-\beta_i))^2
(z-\beta_i)^2}
\nonumber
\\
&+\frac{R''''(\beta_i)}{48 (R''(\beta_i))^2 R'(-\beta_i) (z-\beta_i)^2}
- \frac{(R'''(\beta_i))^2}{48 (R''(\beta_i))^3 R'(-\beta_i) (z-\beta_i)^2}
\Big\}\;.
\end{align}
\end{proposition}
For later purposes, we choose to decompose $ \omega_{1,1}(z) =  \omega^{TR}_{1,1}(z) + \omega^{BTR}_{1,1}(z) $ where the latter one is induced by the additional \textit{blobs} $\phi_{0,2}(u,z)=B(u,-z)$ and $\mathcal{H}_z  \omega_{1,1}(z) $
 \begin{align}
\label{eq.abc}
 \frac{\omega^{BTR}_{1,1}(z)}{dz} =
-\frac{\lambda}{8 (R'(0))^2 z^3}
+\frac{\lambda R''(0)}{16(R'(0))^3z^2}
-\sum_{i=1}^{2d} 
\frac{\lambda}{8 \beta_i^2 R''(\beta_i) R'(-\beta_i)  (z-\beta_i)^2}\; .
\end{align}
 For $d=1$, the complete $\Omega_1^{(1)}$ evaluated at $z=\varepsilon$ gives exactly the expected numbers $ \frac{3^n}{6} ( 2^{2n}-\frac{(2n)!}{n!n!}) $ after performing a power series with computer algebra.

\subsection{The role of the Bergman $\tau$-function}

 We have seen that the Bergman $\tau$-function  did not contribute to $\mathcal{F}^{(1)}$ in the QKM, although the loop insertion operator would act on it (explicit $e$-dependence). But what would be created, if we add the $\tau$-function to $ \mathcal{F}^{(1)}$? Calculating the residue in eq. (\ref{eq.f1gen}) gives the implicit definition of the $\tau$-function via the differential equation
\begin{align*}
  \partial_{b_i} \ln[\tau_B(b_1,...,b_{2d})] =\frac{1}{24}\biggl (\frac{R''''(\beta_i)}{R''(\beta_i)^2}-\frac{R'''(\beta_i)^2}{R''(\beta_i)^3} \biggl )
\end{align*}
with $R(\beta_i)=:b_i$. For $d=1$, the $\tau$ function can be explicitly calculated with $R(z)=z-\frac{\lambda \varrho}{z+\varepsilon}$, $b_{1,2}=-\varepsilon(e,\lambda) \pm 2i \sqrt{\lambda \varrho(e,\lambda)}$. We have the explicit dependence
\begin{align*}
 \varepsilon(e,\lambda) = \frac{1}{6}(4e+\sqrt{4e^2+12\lambda}) \;, \qquad
  \varrho(e,\lambda)= \frac{N}{18 \lambda} (2e\sqrt{4e^2+12\lambda}-4e^2+12\lambda)\;.
\end{align*}
 Then we can solve $\ln[\tau_B(b_1,b_2)]$ by calculating $\frac{1}{24}\big (\frac{R''''(\beta_i)}{R''(\beta_i)^2}-\frac{R'''(\beta_i)^2}{R''(\beta_i)^3} \big ) = \frac{\varepsilon+\beta_i}{16 \lambda \varrho}$ by rewriting this expression as a function of $b_{1,2}$. We have thus shown:
\begin{lemma}
The Bergman $\tau$-function as a multivalued function of the branch points $b_i$ reads for $d=1$ in the QKM:
\begin{align*}
\ln[\tau_B(b_1,b_2)]= \ln\bigg ( \frac{b_1-b_2}{4} \bigg ) =\ln(- i \sqrt{\lambda\varrho}) 
\end{align*}
\end{lemma}
We remark that the the calculation according to eq. (\ref{eq.f1gen}) only contained the standard bidifferential $B(q, \sigma(q))$. In App. \ref{appB} we discuss the role of the additional reflected Bergman kernel that would contribute as $B(q,-\sigma(q))$. The surprising fact about taking this first blob $\phi_{0,2}$ into account is that does not significantly affect the result of $\ln(\tau_B)$. It simply reads $\ln(+ i \sqrt{\lambda\varrho}) $. 

\begin{figure}[h!]
  \centering
    \includegraphics[width= 0.72\textwidth]{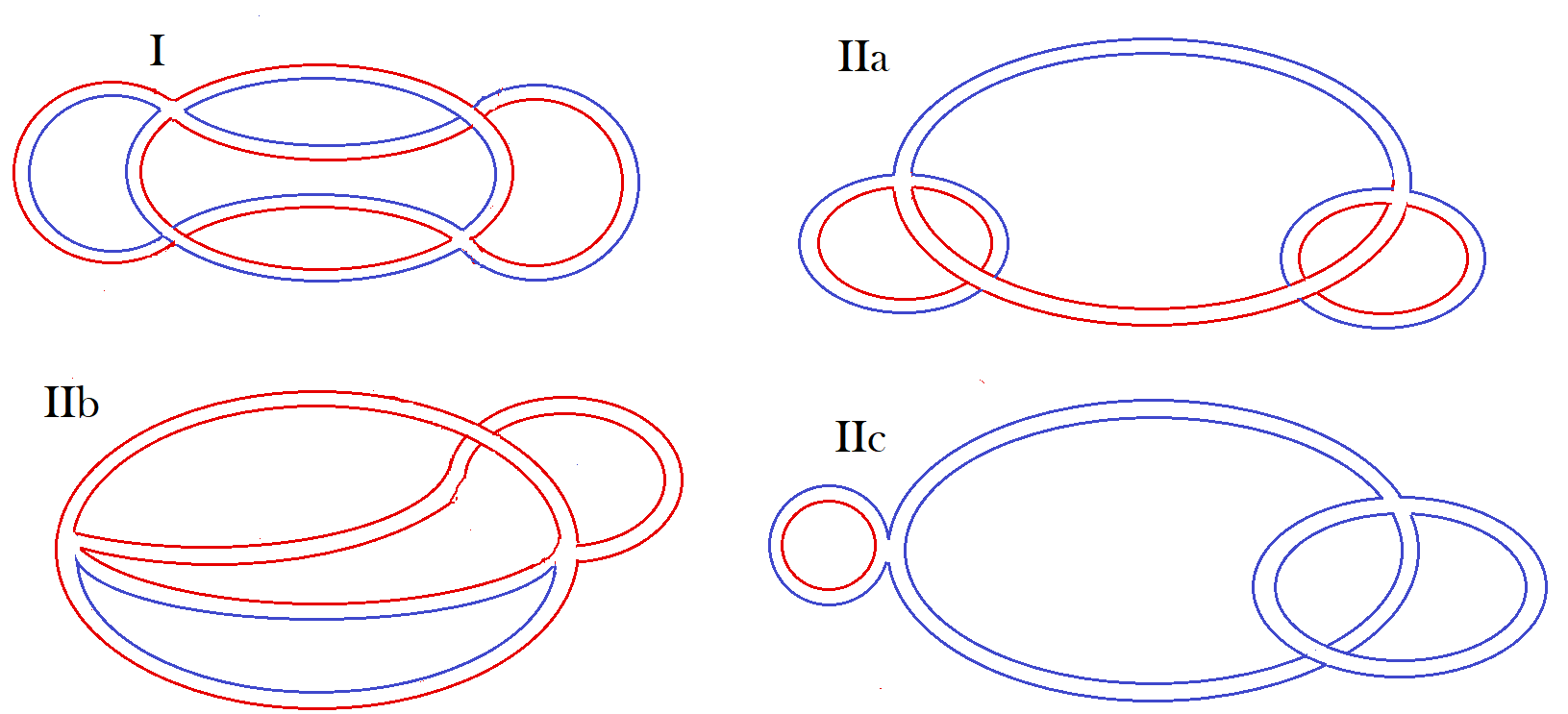} 
    \caption{$\mathcal{O}(\lambda^2)$ contributions to $\mathcal{F}^{(1)}$, ordered in decreasing symmetry. The highly-symmetric graph I carries the biggest possible automorphism group $|\mathrm{Aut}(\Gamma_I)|=8$, followed by $|\mathrm{Aut}(\Gamma_{IIa})|=4$, $|\mathrm{Aut}(\Gamma_{IIb})|=2$ and $|\mathrm{Aut}(\Gamma_{IIc})|=1$, the sum of the reciprocals yields the expected $\frac{15}{8}$. Applying the boundary creation operator gives as numbers of generated graphs the reciprocal of the automorphism group. Each bi-colored ribbon, cut by $\hat T$, contributes to the $g=1$ two-point function, the rest to the 1+1-point function - 15 different graphs in total..
\label{graph2}}
\end{figure}

\begin{figure}[h!]
  \centering
    \includegraphics[width= 0.72\textwidth]{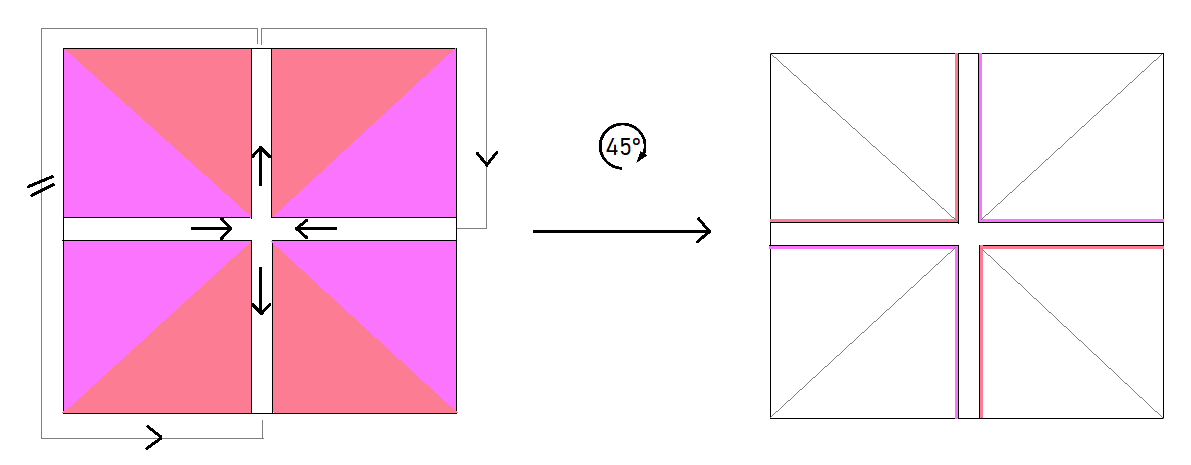} 
    \caption{Create a chequered surfaces by coloring in- and outgoing arrows from a vertex by two colours. Adjacent faces must not have the same colour. Therefore, one cannot glue the square to a torus (no contribution for $\lambda^1$ as observed), whereas all graphs in $\mathcal{F}^{(0)}$ (see \cite{Branahl:2020b} Fig. 3.2) correspond to chequered surfaces. We obtain our picture of bi-coloured ribbon graphs by turning the picture by 45 degrees around. From this viewpoint, a torus cannot be constructed since \textit{different} colours of the ribbon meet during the gluing.
\label{graph3}}
\end{figure}

\begin{figure}[h!]
  \centering
    \includegraphics[width= 0.92\textwidth]{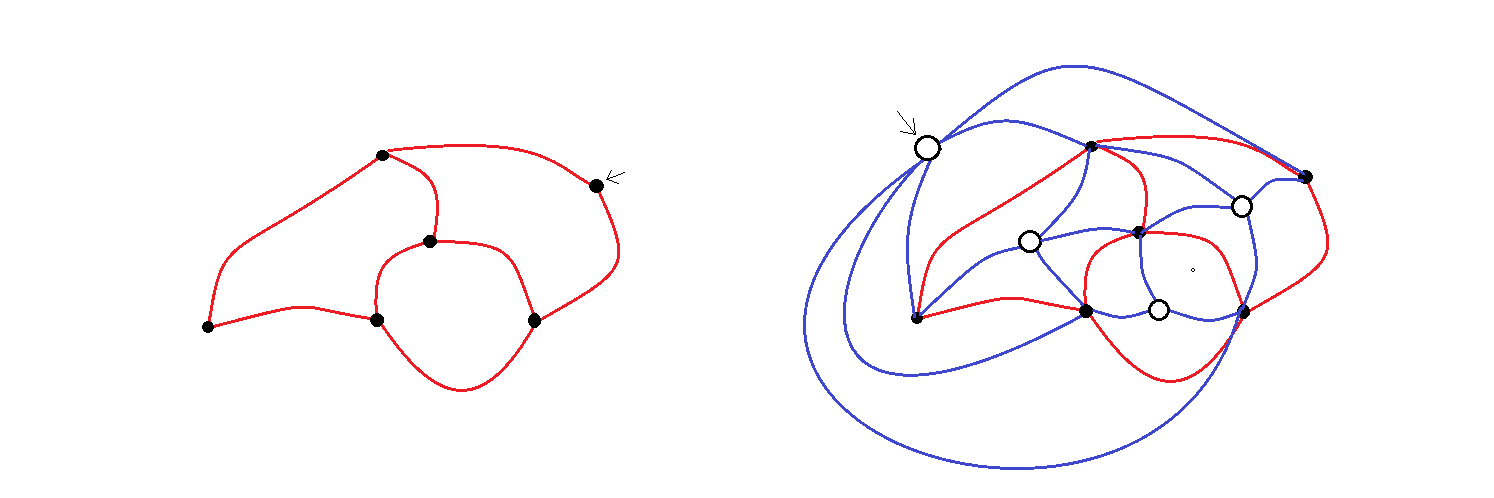} 
    \caption{Consider an arbitrary rooted map with $n$ edges and genus $g$, for instance two quadrangles, one triangle and 5-angle each (left-hand side). If one inserts a new type of vertices (here: white) into every face of the map, there is always the possibilty to draw new edges (blue) such that all the white vertices are connected only with black vertices, the old edges are not intersected and the blue edges form only quadrangles (8). This phenomenon is called \textit{Tutte's bijection} of rooted maps and rooted bipartite quadrangulations. For $g=0$, all quadrangulations are bipartite, for higher genus the bipartite ones are only a small subset. The results also hold if one removes the roots.
\label{tutte}}
\end{figure}

Let us add the $\tau$-function to the previous $\lambda$-expansion. To our surprise, also the coefficients of the expansion of $ -\frac{1}{2} \ln[\tau_B(b_1,b_{2})]  + \mathcal{F}^{(1)}$ in $\lambda$ have a deep meaning:
\begin{proposition}
\label{f1d1tau}
The genus one free energy of the quartic Kontsevich model together with the Bergman $\tau$-function reads in the special case $d=1$:
\begin{align}
\label{f1tau}
 & -\frac{1}{2} \ln[\tau_B(b_1,b_{2})]  + \mathcal{F}^{(1)}  \nonumber  \\
&= -\frac{i\pi}{4}+ \frac{1}{12}\sum_{n=0}^{\infty} \frac{3^{n+1}}{n+1} \sum_{p=0}^n \frac{(2n+2)!}{(n-p)!(n+2+p)!}(1-(-3)^{-p})\frac{\lambda^{n+1}}{(2e)^{2n+1}}
\end{align}
\end{proposition}
We will now show that we have created a primitive of the normalised solution $\Omega^{TR}_{1,1}(\varepsilon)$ with respect to the creation operator ($\frac{i\pi}{4}$ vanishes after deriving the result. The apperanec of an imaginary term already suggests that the $\tau$-function must not play a role in $\mathcal{F}^{(1)}$ due to contradictions to the perturbative expansion.) and give a combinatorial interpretation as well. The coefficients of the series in (\ref{f1tau}), 
\begin{align*}
\bigg \{0,  \frac{1}{2}, \frac{20}{3}, \frac{307}{4}, \frac{4280}{5}, \frac{56914}{6},... \bigg \} \; ,
\end{align*}
 have been investigated in different contexts since the 90's: In \cite{Morris:1990cq}, the author used the \textit{complex matrix model} to count \textit{chequered surfaces} (like on a chess board, adjacent faces must not have the same color). With the same model, also generating functions for counting \textit{links and tangles} were developed, especially beyond $g=1$ \cite{Zinn-Justin:2003ecd}. To illustrate the difference when one allows the surplus term $\ln(\tau_B)$, we look at Fig. \ref{graph2} containing the $\mathcal{O}(\lambda^2)$ contributions to $\mathcal{F}^{(1)}$.
The only graph that has got bi-colored ribbons everywhere is the first one. We motivate briefly that the explanation via bi-colored ribbons is equivalent to the counting problem of chequered surfaces - they are complementary to each other.  Ribbon graphs are created in the quartic Kontsevich model as by contracting the indices representing propagators (Gaußian term in the partition function) and tetravalent vertices (quartic potential).
\smallskip
 \begin{align*}
\begin{picture}(30,15)
  \put(0,5){\line(1,0){30}}
  \put(0,9){\line(1,0){30}}
  \put(2,-3){\mbox{\scriptsize$j$}}
  \put(2,11){\mbox{\scriptsize$i$}}
  \put(28,-3){\mbox{\scriptsize$l$}}
  \put(28,11){\mbox{\scriptsize$k$}}
\end{picture} \quad =N\delta_{jl}\delta_{ik}, \qquad \quad  \begin{picture}(24,19)
  \put(0,6){\line(1,0){10}}
  \put(0,10){\line(1,0){10}}
  \put(24,6){\line(-1,0){10}}
  \put(24,10){\line(-1,0){10}}
  \put(10,-4){\line(0,1){10}}
  \put(14,-4){\line(0,1){10}}
  \put(10,20){\line(0,-1){10}}
  \put(14,20){\line(0,-1){10}}
  \put(5,-9.5){\mbox{\scriptsize$k$}}
  \put(15,-9.5){\mbox{\scriptsize$l$}}
  \put(14,21.5){\mbox{\scriptsize$p$}}
  \put(3,21.5){\mbox{\scriptsize$q$}}
  \put(-10,0){\mbox{\scriptsize$j$}}
  \put(-10,8){\mbox{\scriptsize$i$}}
  \put(26,0){\mbox{\scriptsize$m$}}
  \put(26,8){\mbox{\scriptsize$n$}}
\end{picture} \qquad  =\lambda \delta_{qi}\delta_{jk}\delta_{lm}\delta_{np}
\end{align*}
\\ \\ 
Complex matrices instead demand additional information for the propagator responsible for $M$ and $M^{\dagger}$, here denoted by an arrow:
\\ 
 \begin{align*}
\begin{picture}(30,15)
  \put(0,5){\line(1,0){30}}
  \put(0,9){\line(1,0){30}}
\put(8,7){\vector(1,0){15}}
  \put(2,-3){\mbox{\scriptsize$j$}}
  \put(2,11){\mbox{\scriptsize$i$}}
  \put(28,-3){\mbox{\scriptsize$l$}}
  \put(28,11){\mbox{\scriptsize$k$}}
  \put(37,4){\mbox{\scriptsize$M$}}
  \put(-17,4){\mbox{\scriptsize$M^{\dagger}$}}
\end{picture}   \qquad \qquad  \qquad  \begin{picture}(24,19)
  \put(0,6){\line(1,0){10}}
  \put(0,10){\line(1,0){10}}
  \put(24,6){\line(-1,0){10}}
  \put(24,10){\line(-1,0){10}}
  \put(10,-4){\line(0,1){10}}
  \put(14,-4){\line(0,1){10}}
  \put(10,20){\line(0,-1){10}}
  \put(14,20){\line(0,-1){10}}
  \put(5,-9.5){\mbox{\scriptsize$k$}}
  \put(15,-9.5){\mbox{\scriptsize$l$}}
  \put(14,21.5){\mbox{\scriptsize$p$}}
  \put(3,21.5){\mbox{\scriptsize$q$}}
  \put(-10,0){\mbox{\scriptsize$j$}}
  \put(-10,8){\mbox{\scriptsize$i$}}
  \put(26,0){\mbox{\scriptsize$m$}}
  \put(26,8){\mbox{\scriptsize$n$}}
\put(16,8){\vector(1,0){9}}
\put(9,8){\vector(-1,0){9}}
\put(12,22){\vector(0,-1){9}}
\put(12,-6){\vector(0,1){9}}
\end{picture}  
\end{align*}
\\ 
The tetravalent vertex is created by $\frac{1}{4}MM^{\dagger}MM^{\dagger}$. In Fig. \ref{graph3} the complementarity between chequered surfaces and bi-colored ribbons is shown.  
At  $\mathcal{O}(\lambda)$ both approaches forbid the construction of any contribution, one has to continue at  $\mathcal{O}(\lambda^2)$. Only the highly symmetric structure of graph \textbf{I} allows to draw the arrows as required by complex matrices. In \cite{Eynard:2005} the natural equivalence to the Hermitian two-matrix model is mathematically shown. We recall the key message that
\begin{align*}
\int_{(H_N\times H_N)(\Gamma)}dM_1dM_2e^{-\mathrm{Tr}[V_1(M_1)+V_2(M_2)+M_1M_2]}= \int_{GL_N(\mathbb{C})}dMe^{-\mathrm{Tr}[V_1(M)+V_2(M^{\dagger})+MM^{\dagger}]}
\end{align*}
holds - the complex matrix model as an analytic continuation of the two-matrix model for some contour $\Gamma$. The two-matrix model exactly obeys topological recursion in case of non-mixed boundaries and thus gives a natural explanation for the observed behaviour.\\ \\ 
In order to formulate the observation for the duals of the ribbon graphs, maps, let us remind the reader of the concept of bipartite quadrangulations. W. Tutte's seminal discovery was that rooted maps are always in bijection with bipartite quadrangulations \cite{Tuttbij}, which are constructed in the way shown in Fig. \ref{tutte}.

\begin{itemize}
\item $\mathcal{F}^{(1)}$  is a generating function for the numbers of quadrangulations of the torus with $k$ faces - the same result as for the known result of the Hermitian 1MM, but \textit{with} its Bergman $\tau$-function.
\item $ -\frac{1}{2} \ln[\tau_B(b_1,b_{2})]  +\mathcal{F}^{(1)}$ is a generating function for the numbers  of \textit{bipartite} quadrangulations of the torus with $k$ faces.
\end{itemize}
Let us apply $\hat T$ on the power series of $ -\frac{1}{2} \ln[\tau_B(b_1,b_{2})]  +\mathcal{F}^{(1)}$. Then we obtain the special numbers $\{1,20,307,4280,...\}$ being the expansion coefficients of $\Omega_{1,1}^{TR}(\varepsilon_1)$ known since the important combinatorial works of Bender and Canfield \cite{Bender}.  We summarize again: 
\begin{itemize}
\item  $\Omega_{1,1}(\varepsilon_1)$ of the QKM is a generating function for rooted quadrangulations with $k$ faces of the torus 
\item Omit any BTR contributions, consider only the normalised solution: $\Omega_{1,1}^{TR}(\varepsilon_1)$ of the QKM is a generating function for rooted \textit{bipartite} quadrangulations with $k$ faces of the torus.   
\end{itemize}
 We finally give an overview of all treated ribbon graph contributions (remember $\Omega_{1,1}(\varepsilon)=\mathcal{G}^{(0)}(\varepsilon|\varepsilon)+\mathcal{G}^{(1)}(\varepsilon,\varepsilon)$): 
\\ \\ 
\centerline{\begin{tabular}[h!b]{|c|c|c|c||c|}
\hline
Order &$\mathcal{G}^{(0)}(\varepsilon|\varepsilon)$&$\mathcal{G}^{(1)}(\varepsilon,\varepsilon)$&$\Omega_{1,1}(\varepsilon)$ &$\Omega_{1,1}^{TR}(\varepsilon)$ \\
\hline
$\lambda^1$ & 1 & 0 & 1 & 0 \\
\hline
$\lambda^2$ & 9 & 6 & 15 & 1 \\
\hline
$\lambda^3$ & 81 & 117 & 198 & 20\\
\hline
$\lambda^4$ & 756 & 1755 & 2511 & 307 \\
\hline
$\lambda^5$ & 7290 & 23976 & 31266 & 4280  \\
\hline
$\lambda^6$ & 72171 & 313227 & 385398  & 56914 \\
\hline
\end{tabular}}
\\ \\ 
coinciding at $\mathcal{O}(\lambda^2)$ with the automorphism groups explained in Fig. \ref{graph2} \footnote{These numbers are in agreement with the "supporting data for quadrangulations" in \cite{Borot:2017agy}}. We remarked in previous works \cite{Branahl:2020b} that the correlation functions are generators for \textit{fully simple} quadrangulations in the sense of Borot and Garcia Failde (\cite{Borot:2017agy} and \cite{Borot:2021eif}). These maps are obtained after exchanging $x$ and $y$ in the spectral curve of the 1MM. In the table from above we recover in a completely different and natural way their relation that ordinary tori (here: $\Omega_{1,1}(\varepsilon)$) with boundary length two of the 1MM decay into fully simple tori and fully simple cylinders with boundary length (1,1), namely $\mathcal{G}^{(0)}(\varepsilon|\varepsilon)+\mathcal{G}^{(1)}(\varepsilon,\varepsilon)$. This holds to all $g$ extending the result in \cite{Borot:2017agy}. \\ \\ 
 Based on the upper explanation we can formulate the justified conjecture that the results must hold for any $g$. We can continue with $g=2$ and conjecture \footnote{The coefficients of the power series for higher $g$ in $\Omega_{g,1}^{TR}$ can be read off from OEIS: A238396, columnwise}:
\begin{align*}
&\Omega_{2,1}(\varepsilon)=45\lambda^2+2007\lambda^3+56646\lambda^4+1290087\lambda^5+... \\ 
&\Omega_{2,1}^{TR}(\varepsilon)= 21\lambda^4+966\lambda^5+27954\lambda^6+650076 \lambda^7+...
\end{align*}
This was shown to be correct with computer algebra, we assume that the structure goes on for all higher genera. The authors of \cite{Carrell:2014} give an astonishing recursion formula for the number $\mathcal{Q}_k^{(g)}$ of \textbf{bipartite} rooted quadrangulations with $k$ faces of a genus-$g$ surface: Our analysis gives a completely new interpretation of the blobs in the QKM:
\begin{conjecture}
\label{bipconj}
Let the external matrix in the quartic Kontsevich model be a scalar multiple of the identity matrix, $E=e \cdot \mathrm{id}_N$. The topological recursion formula for the spectral curve $(x:\hat{\mathbb{C}}\to \hat{\mathbb{C}},
\omega_{0,1}=ydx,B)$ with 
\begin{align*}
&x(z) = z- \frac{\lambda}{N}\frac{\varrho}{z+\varepsilon} \qquad y(z)=-x(-z)  \qquad B(z_1,z_2)=\frac{dz_1dz_2}{(z_1-z_2)^2}\\
& \varepsilon(e,\lambda) = \frac{1}{6}(4e+\sqrt{4e^2+12\lambda}) \;, \qquad
  \varrho(e,\lambda)= \frac{N}{18 \lambda} (2e\sqrt{4e^2+12\lambda}-4e^2+12\lambda)
\end{align*}
gives generating functions for the number $\mathcal{Q}_k^{(g)}$ of \textbf{bipartite} rooted quadrangulations with $k$ faces of a genus-$g$ surface:
\begin{align*}
&\frac{\omega^{TR}_{g,1}(\varepsilon) }{dR(\varepsilon)}= \frac{1}{R'(\varepsilon)} \Res_{q \to - \varepsilon \pm i \sqrt{\lambda \varrho}} \big ( K_{\pm}(q,\sigma(q),\varepsilon) \cdot \omega_{g-1,2}(q,\sigma(q)) \big )  = \sum_{k=0}^{\infty} \mathcal{Q}_k^{(g)} \frac{(-\lambda)^k}{(2e)^k}
\end{align*}
for $g>0$ with the global Galois involution $\sigma(q) = \varepsilon+ \frac{\lambda}{N}\frac{\varrho}{q+\varepsilon} $.
\end{conjecture}
For $g=0$, all quadrangulations of the sphere are automatically bipartite in accordance with the fact that $\omega_{0,1}$ is unaffected by blobbed topological recursion. To the best of our knowledge, a concrete formula in the framework of topological recursion for bipartite quadrangulations of a genus-$g$ surface is not given in the literature. Similar to the appearance of blobbed topological recursion in the multi-trace case in the Hermitian 1MM, if one would like to generate \textit{stuffed maps} \cite{Borot:2013fla}, having the ordinary maps as a subset, also here the blobs are responsible for the generation of a larger class of maps: Also the non-bipartite quadrangulations, a class of maps with less symmetry, are generated in BTR. 

Not a conjecture, but a statement is that the holomorphic additions $\mathcal{H} \omega_{g,n}$ have the following interpretation in terms of enumerative geometry: The complete results for $\omega_{g,1}$ are generating functions for rooted quadrangulations of a genus-$g$ surface, bipartite and non-bipartite ones, the subset of the first is created by the normalised solutions in the sense of \cite{Borot:2015hna}. Based on the one-to-one correspondence to the partition function of the Hermitian one-matrix model and therefore equivalent graph expansions of the free energies $\mathcal{F}^{(g)}$ together with the aforementioned equivalence of the boundary creation operator for the creation of $n=1$ boundary, there is nothing to prove. The conjecture above might be explained by an equivalence of our spectral curve and a special case of the spectral curve in the two-matrix model, which has to be figured out.

 The characterisation of the Feynman diagrams generated by $\omega_{g,n}^{TR}$ for $n>1$ is work in progress.

\section{Generalisation to Arbitrary Simple Ramifications}
\label{ch:proof}
 Building on the combinatorial construction of the genus-one free energy for two simple ramification points, we now aim to prove the result in full generality, that is arbitrary $d$. Remarkably, the structure stays the same apart from a compensation term that we will comment on at the end of the proof.

\subsection{The Formal Action of the Creation Operator}
  Starting from the definition of the creation operator, we seem at first sight to deal with a common derivative with respect to an eigenvalue of the external matrix $E$. However, when we aim to apply the operator to $\omega_{g,n}(z_1,...,z_n)$, we are faced with two obstacles: The first challenge occurs after the complexification of $\Omega_{q_1,...,q_n}^{(g)}$ and the variable transform using the function $R(z)$: Any variable $z_i$ has to be seen as a pull-back from the $R$-plane, $z_i=R^{-1}(x_i)$, where $R^{-1}$ is dependent on the eigenvalues $e_b$. Only $x_i$ is in the end untouched by the creation operator as long as its explicit value is not $e_b$-dependent. The second challenge is the complexity of $R(z)$, since beyond $d=1$ the constituents of $R(z)$, namely $\{\varepsilon_k\}_{k=1,...d}$ and $\{\varrho_k\}_{k=1,...d}$ with $\varrho_k=\frac{r_k}{R'(\varepsilon_k)}$, are only implicitly given as a functions like $\varepsilon_k(e_1,...,e_d)$ with no global representation.

It is necessary to grasp the creation operator acting on the complexified $\omega_{g,n}(z_1,...,z_n)$ as a formal derivation, in particular to give a meaning to $\hat T_b z_i$, which cannot be formulated with an ordinary derivative in the $z_i$-plane. We will distinguish the ordinary derivative $\partial_{e_b}$ from the insertion operator containing the derivation $\partial^{int}_{e_b}$, where \textit{int} stands for \textit{internal} (inner function). 
\begin{definition}
	We define $\partial^{int}_{e_b}$ by the following properties
	\begin{enumerate}
		\item $\partial^{int}_{e_b}$ acts as a formal derivative and fulfils the Leibniz rule   as well as the chain rule:
		\begin{align*}
			\partial^{int}_{e_b} (fg) = f \partial^{int}_{e_b} g + g \partial^{int}_{e_b} f \\
			\partial^{int}_{e_b}(f(z)) = (\partial^{int}_{e_b} f) (z)+ f'(z) \partial^{int}_{e_b} z
		\end{align*}
		\item Set $\partial^{int}_{e_b} R(z) = 0$ 
		\item Property of boundary creation: 
		\begin{align*}
			-\frac{N\lambda}{r_b} \partial^{int}_{e_b} \omega_{g,n}(z_1,...,z_n) dR(\varepsilon_b)= \omega_{g,n+1}(z_1,...,z_n,\varepsilon_b)
		\end{align*} 
	\end{enumerate}
\end{definition}
Property (b) directly follows from $R(z_i)=R(R^{-1}(x_i)=x_i$ which is unaffected by the action of $-\frac{N}{r_b} \partial^{int}_{e_b} $.
The most important observation to make the proof for $\mathcal{F}^{(1)}$ work is the transition $\partial^{int}_{e_b}$ to $\partial_{e_b}$ for the free energy (no dependence on complex variables anymore). This proof will contain a lemma that emphasises the differences between the formal creation operator and $\partial_{e_b}$ highlighting once more the importance of the symmetry $z \leftrightarrow -z$:
\begin{lemma}
\label{lem:1}
\begin{align*}
 \partial_{e_b}R(-z)+\frac{R'(-z)}{R'(z)} \partial_{e_b}R(z)= - \frac{r_b}{N} \frac{1}{R'(z)R'(\varepsilon_b)} \biggl ( \frac{1}{(z-\varepsilon_b)^2}+\frac{1}{(z+\varepsilon_b)^2} \biggl )
\end{align*}
where $\frac{1}{(z-\varepsilon_b)^2}+\frac{1}{(z+\varepsilon_b)^2}= \frac{\omega_{0,2}(z,\varepsilon_b)}{dz\, d\varepsilon_{b}}$.
\end{lemma}

\textit{Proof.}  
  We know from the definition of the insertion operator $\partial^{int}_{e_b} R(-z)=- \frac{r_b}{N} \frac{\omega_{0,2}(z,\varepsilon_b)}{dR(z)\,dR(\varepsilon_b)}   $. We can equate this result with an ordinary derivative due to $\partial^{int}_{e_b} R(-z) =  \partial_{e_b} R(-z)-R'(-z) \partial^{int}_{e_b} z$ for $z \notin\{\pm \varepsilon_k, \pm \beta_i \}$. Furthermore, we have claimed $\partial^{int}_{e_b} R(z)=0$, which can be again rewritten using $\partial^{int}_{e_b} R(z)= \partial_{e_b} R(z)+R'(z) \partial^{int}_{e_b} z$. These two conditions give the possibility to
 eliminate $\partial^{int}_{e_b} z$ in order to obtain the lemma as claimed. \qed \\ \\ 

Interestingly there is, despite the symmetry of $z \to - z$ in the two summands in  $\omega_{0,2}$, no one-to-one correspondence to the two summands on the left-hand side. Instead there is a non-trivial entanglement of both. We conclude with the natural consequence of this Lemma for the ordinary derivative for general $\Omega_{g,n}$:
\begin{corollary}
In terms of the ordinary derivative, the boundary creation can be carried out as follows:
\begin{align*}
  - \frac{r_b}{N}\Omega_{g,n+1}(z_1,...,z_n,\varepsilon_b)= \partial_{e_b}\Omega_{g,n}(z_1,...,z_n)-\sum_{i=1}^n  \frac{\partial}{\partial R(z_i) } \Omega_{g,n}(z_1,...,z_n)  \partial_{e_b}R(z_i)
\end{align*}
\end{corollary}
 As the next step, we try to give a meaning to the internal derivative acting on an arbitrary $z$:

\begin{lemma}
\label{lem:2}
The internal derivative $\frac{\partial^{int}}{\partial e_b}$ acts on an arbitrary variable $z \notin \{\beta_i, \varepsilon_k\}$ as follows:
\begin{align*}
&-\frac{N}{\lambda}\frac{R'(\varepsilon_b)}{r_b}\frac{\partial^{int} z}{\partial e_b}=\frac{1}{R'(z)R'(-z)(z+\varepsilon_b)^2}\\&+\sum_{i=1}^{2d}\frac{1}{(\beta_i-\varepsilon_b)^2}\frac{1}{R'(-\beta_i)R''(\beta_i)}\bigg(\frac{1}{(z-\beta_i)}+\frac{1}{(z+\beta_i)}\bigg)
\end{align*}
\end{lemma}

\textit{Proof.} 
We start by writing out the steps of the proof of Lemma \ref{lem:1} explicitly:

\begin{align*}\nonumber
	&\hat T_b R(-z)=-\frac{N}{r_b}\frac{\partial^{int} z}{\partial e_b}\bigg(1+\frac{\lambda}{N}\sum_n\frac{r_n}{R'(\varepsilon_n)(\varepsilon_n-z)^2}\bigg)\\
&+\frac{\lambda}{r_b}\sum_n
	\frac{r_n}{(R'(\varepsilon_n))^2(\varepsilon_n-z)}\frac{\partial R'(\varepsilon_n)}{\partial e_b}\\
	&+\frac{\lambda}{r_b}\sum_n
	\frac{r_n}{R'(\varepsilon_n)(\varepsilon_n-z)^2}\frac{\partial \varepsilon_n}{\partial e_b}.
\end{align*}
 The definition of $\frac{\partial^{int} }{\partial e_b}$ (recall $\hat T_b R(z)=0$) gives the relation
\begin{align}
	 0=&\,R'(z)\frac{\partial^{int} z}{\partial e_b} +\frac{\lambda}{N}\sum_n \frac{r_n}{R'(z)R'(\varepsilon_n)(\varepsilon_n+z)^2}
	\frac{\partial \varepsilon_n}{\partial e_b}
	+\frac{\lambda}{N}\sum_n \frac{r_n}{R'(z)R'(\varepsilon_n)^2(\varepsilon_n+z)}
	\frac{\partial R'(\varepsilon_n)}{\partial e_b} 
\end{align}
Combining these two equations by eliminating once again $\frac{\partial^{int} }{\partial e_b}z$ yields

		\begin{align}\nonumber
		&\frac{R'(-z)}{r_b}\bigg(\sum_k \frac{r_k}{R'(\varepsilon_k)(\varepsilon_k+z)^2}
		\frac{\partial \varepsilon_k}{\partial e_b}
		+\sum_k \frac{r_k}{R'(\varepsilon_k)^2(\varepsilon_k+z)}
		\frac{\partial R'(\varepsilon_k)}{\partial e_b}\bigg)\\\nonumber
		&+\frac{R'(z)}{r_b}\bigg( \frac{1}{r_b}\sum_k
		\frac{r_k}{R'(\varepsilon_k)(\varepsilon_k-z)^2}\frac{\partial \varepsilon_k}{\partial e_b} + \sum_k
		\frac{r_k}{(R'(\varepsilon_k))^2(\varepsilon_k-z)}\frac{\partial R'(\varepsilon_k)}{\partial e_b}
		\bigg)\\
		=&\frac{1}{R'(\varepsilon_b)(z-\varepsilon_b)^2}+\frac{1}{R'(\varepsilon_b)(z+\varepsilon_b)^2}.\label{RE}
		\end{align}

We evaluate \eqref{RE} at $z=-\beta_i$ with $R'(\beta_i)=0$:
\begin{align}\nonumber
		&\frac{1}{r_b}\sum_k
		\frac{r_k}{(R'(\varepsilon_k))^2(\varepsilon_k+\beta_i)}\frac{\partial R'(\varepsilon_k)}{\partial e_b}
		+\frac{1}{r_b}\sum_k
		\frac{r_k}{R'(\varepsilon_k)(\varepsilon_k+\beta_i)^2}\frac{\partial \varepsilon_k}{\partial e_b}\\\label{betaid}
		=&\frac{1}{R'(-\beta_i)R'(\varepsilon_b)(\beta_i+\varepsilon_b)^2}+\frac{1}{R'(-\beta_i)R'(\varepsilon_b)(\beta_i-\varepsilon_b)^2}.
\end{align}
We rewrite the identity \eqref{betaid} to
\begin{align}\nonumber
		&\sum_k
		\frac{r_k(\varepsilon_k+\beta_i)\prod_{n\neq k}(\varepsilon_n+\beta_i)^2}{(R'(\varepsilon_k))^2}\frac{\partial R'(\varepsilon_k)}{\partial e_b}
		+\sum_k
		\frac{r_k\prod_{n\neq k}(\varepsilon_n+\beta_i)^2}{R'(\varepsilon_k)}\frac{\partial \varepsilon_k}{\partial e_b}\\\label{betaid1}
		=&\frac{r_b\prod_n(\varepsilon_n+\beta_i)^2}{R'(-\beta_i)R'(\varepsilon_b)}\bigg(\frac{1}{(\beta_i+\varepsilon_b)^2}+\frac{1}{(\beta_i-\varepsilon_b)^2}\bigg).
\end{align}
and in the same manner:
\begin{align*}
&\sum_k\frac{r_k}{(R'(\varepsilon_k))^2(\varepsilon_k+z)}\frac{\partial R'(\varepsilon_k)}{\partial e_b}
		+\frac{1}{r_b}\sum_k
		\frac{r_k}{R'(\varepsilon_k)(\varepsilon_k+z)^2}\frac{\partial \varepsilon_k}{\partial e_b} \\
=&\frac{1}{\prod_{n}(z+\varepsilon_n)^2}\bigg(\sum_k
		\frac{r_k(\varepsilon_k+z)\prod_{n\neq k}(\varepsilon_n+z)^2}{(R'(\varepsilon_k))^2}\frac{\partial R'(\varepsilon_k)}{\partial e_b}
		+\frac{1}{r_b}\sum_k
		\frac{r_k\prod_{n\neq k}(\varepsilon_n+z)^2}{R'(\varepsilon_k)}\frac{\partial \varepsilon_k}{\partial e_b} \bigg),
\end{align*}
This gives rise to a polynomial of degree $2d-1$ within the parentheses from which we know $2d$ points given by \eqref{betaid1}. Now recall the Lagrange interpolation formula: Let $f$ be a polynomial of degree $d-1\geq 0$ and 
$x_1,...,x_d$ be pairwise distinct complex numbers.
Then, for all $x\in \mathbb{C}$,
\begin{align*}
		f(x)=L(x)\sum_{j=1}^d\frac{f_j}{(x-x_j)L'(x_j)},\qquad 
\text{where } L(x)=\prod_{j=1}^d(x-x_j) \text{ and } f_j=f(x_j).
\end{align*}
Using 
\begin{align*}
R'(z)=\frac{\prod_{i=1}^{2d}(z-\beta_i)}{\prod_{n=1}^d (z+\varepsilon_n)^2}.
\end{align*}
valid due to the fundamental theorem of algebra, the interpolation formula yields
\begin{align*}
&\sum_k\frac{r_k}{(R'(\varepsilon_k))^2(\varepsilon_k+z)}\frac{\partial R'(\varepsilon_k)}{\partial e_b}
		+\sum_k
		\frac{r_k}{R'(\varepsilon_k)(\varepsilon_k+z)^2}\frac{\partial \varepsilon_k}{\partial e_b} \\
=&\frac{r_b}{R'(\varepsilon_b)}\frac{\prod_{k=1}^{2d}(z-\beta_k)}{\prod_{n}(z+\varepsilon_n)^2}\sum_{i=1}^{2d}\prod_n(\varepsilon_n+\beta_i)^2\bigg(\frac{1}{(\beta_i+\varepsilon_b)^2}+\frac{1}{(\beta_i-\varepsilon_b)^2}\bigg)\\&\frac{1}{R'(-\beta_i)(z-\beta_i)\prod_{k\neq i}^{2d} (\beta_i-\beta_k)}\\
=&\frac{r_b}{R'(\varepsilon_b)}R'(z)\sum_{i=1}^{2d}\bigg(\frac{1}{(\beta_i+\varepsilon_b)^2}+\frac{1}{(\beta_i-\varepsilon_b)^2}\bigg)\frac{1}{R'(-\beta_i)(z-\beta_i)R''(\beta_i)}
\end{align*} 
Compare the first line with the results of the proof of Lemma \ref{lem:1}. We directly read off
\begin{align*}
\frac{R'(\varepsilon_b)}{r_b}\frac{\partial^{int} z}{\partial e_b}=-\frac{\lambda}{N}\sum_{i=1}^{2d}\bigg(\frac{1}{(\beta_i+\varepsilon_b)^2}+\frac{1}{(\beta_i-\varepsilon_b)^2}\bigg)\frac{1}{(z-\beta_i)R'(-\beta_i)R''(\beta_i)}.
\end{align*}
again valid for $z\neq \varepsilon_b,\beta_i$. Next, we perform a partial fraction decomposition:
\begin{align}
\label{pfe0}
	\frac{1}{R'(z)R'(-z)(z+w)^2}=&\frac{1}{R'(w)R'(-w)(z+w)^2}+\frac{\frac{R''(w)}{R'(w)}-\frac{R''(-w)}{R'(-w)}}{R'(w)R'(-w)(z+w)}\nonumber \\+&\sum_i\frac{1}{R''(\beta_i)R'(-\beta_i)}\bigg(\frac{1}{(z-\beta_i)(w+\beta_i)^2}-\frac{1}{(z+\beta_i)(w-\beta_i)^2}\bigg)
\end{align}
Insert $w=\varepsilon_b$ (the first two terms vanish) and combine the result with the preliminary formula for $\frac{\partial^{int} z}{\partial e_b}$ finally giving a (for later purposes) suitable representation claimed in the lemma. \qed \\ \\ 
An easy application of Lemma \ref{lem:2} consists of the following calculation:
\begin{proposition}
The internal derivative $\frac{\partial^{int}}{\partial e_b}$ creates another boundary of the $\Omega_{g,n}$ obeying blobbed topological recursion:
\begin{align*}- \frac{N}{r_b}\frac{\partial^{int}}{\partial e_b}  \frac{1}{R'(v)R'(z)} \biggl ( \frac{1}{(v-z)^2} + \frac{1}{(v+z)^2} \biggl ) =\Omega_{0,3}(z,v,\varepsilon_b)
\end{align*}
with 
\begin{align*}
&\Omega_{0,3}(z,v,\varepsilon_b) =\frac{\partial^3}{\partial R(\varepsilon_b)\partial R(z)\partial R(v)}\bigg\{\frac{ Q(v;z)}{R'(z)R'(-z)(z+\varepsilon_b)} + z \leftrightarrow v \\
&+\sum_{i=1}^{2d}\frac{1}{\varepsilon_b-\beta_i}\frac{Q(z;\beta_i)Q(v;\beta_i)}{R'(-\beta_i)R''(\beta_i)}\bigg\}
\end{align*}
and $Q(a;b):=\frac{1}{a+b}+\frac{1}{a-b}$.
\end{proposition}
\textit{Proof.} We write 
\begin{align*}
-\frac{N}{r_b} \frac{\partial^{int}}{\partial e_b}  \frac{1}{R'(v)R'(z)} \biggl ( \frac{1}{(v-z)^2} + \frac{1}{(v+z)^2} \biggl ) 
=\frac{N}{r_b} \frac{\partial^{int}}{\partial e_b} \frac{\partial^2}{\partial R(v)\partial R(z)}\ln \left(\frac{v+z}{v-z}\right)
\end{align*}
in order to commute derivatives and creation operator. $\Omega_{0,3}(z,v,\varepsilon_b)$ can be easily reconstructed by knowing the action of $\frac{\partial^{int}}{\partial e_b} $ on the variables $z$ and $v$:
\begin{align*} 
&\frac{N}{r_b}\frac{\partial^2}{\partial R(z)\partial R(v)}\frac{\partial }{\partial e_b}\ln \left(\frac{v+z}{v-z}\right)=\frac{N}{r_b}\frac{\partial^2}{\partial R(z)\partial R(v)}\bigg ( \frac{\partial^{int}z}{\partial e_b} Q(v,z) + z \leftrightarrow v \bigg )
\end{align*}
Lemma \ref{lem:2} can be rewritten (see proof) as 
\begin{align*}
\frac{N}{\lambda r_b} \frac{\partial^{int} z}{\partial e_b}=\frac{\partial}{\partial R(\varepsilon_b)} \biggl [\frac{1}{R'(z)R'(-z)(z+\varepsilon_b)}+\sum_{i=1}^{2d}\frac{Q(z,\beta_i)}{(\beta_i-\varepsilon_b)}\frac{1}{R'(-\beta_i)R''(\beta_i)}  \biggl ]
\end{align*}
Finally, one only has to check the simple equation $Q(v,z)Q(z,\beta_i)+Q(z,v)Q(v,\beta_i)=Q(z,\beta_i)Q(v,\beta_i)$. \qed \\ \\ 
 We refer to the result of $\Omega_{0,3}(z,v,u)$ after analytic continuation of $u=\varepsilon_b$ in Chapter 5 of \cite{Branahl:2020yru} being exactly what was produced in the proposition from above. Moreover, we can illustrate with perturbation theory in the combinatorial limit of $d=1$ that the creation operator cannot be an ordinary derivative if it does not act on the free energies/closed ribbon graphs. The first orders in a perturbative series in $\lambda$ read for 
\begin{align*}
\Omega_2^{(0)} (\varepsilon,\varepsilon) = \frac{1}{(2e)^2} + \frac{7(-\lambda)}{(2e)^4} +\frac{58(-\lambda)^2}{(2e)^6} +\frac{522(-\lambda)^3}{(2e)^8}+\frac{4941(-\lambda)^4}{(2e)^{10}} + \mathcal{O}(\lambda^5)\\
\Omega_3^{(0)} (\varepsilon,\varepsilon,\varepsilon) = \frac{12(-\lambda)}{(2e)^5}+\frac{240(-\lambda)^2}{(2e)^{7}}+\frac{3628(-\lambda)^3}{(2e)^{9}} + \frac{49464(-\lambda)^4}{(2e)^{11}}+ \mathcal{O}(\lambda^5)
\end{align*}
One directly observes that the coefficients from above cannot be generated with the use of an ordinary derivative $\partial_e$ as it was possible for $\mathcal{F}^{(1)}$.

\subsection{Proof of the Main Theorem}
Having in mind the Lemmata \ref{lem:1} and \ref{lem:2}, we are now ready to carry out the proof of the main theorem in full generality:
\begin{theorem}
The free energy $\mathcal{F}^{(1)}$ reads for the quartic Kontsevich model:
 \begin{align*}
\mathcal{F}^{(1)}=  \frac{\mathfrak{R}_{\neq}}{24} -\frac{1}{24} \ln \biggl ( R'(0) \prod_{i=1}^{2d} R'(-\beta_i) \biggl )
\end{align*}
$\mathfrak{R}_{\neq}$ vanishes for $d=1$ and can be given approximately in terms of the eigenvalues of the external matrix at $\mathcal{O}(\lambda)$:
\begin{align*}
\mathfrak{R}_{\neq} = -\frac{\lambda}{N} \sum_{k,l, k\neq l}^d \frac{r_kr_l}{(e_k+e_l)^2} + \mathcal{O}(\lambda^2)
\end{align*}
It is a primitive of the complicated term
\begin{align}\label{simpl}
&  \hat T_b \frac{\mathfrak{R}_{\neq}}{24} = \sum_i \frac{1}{24 R'(-\beta_i)R''(\beta_i)}\bigg[ - \frac{1}{(\beta_i-\varepsilon_b)^2} \biggl ( \sum_j\frac{1}{2(\beta_i+\beta_j)^2} \nonumber \\ 
&+\sum_k\frac{4}{(\varepsilon_k+\beta_i)^2} 
+  \sum_{j \neq i} \frac{1}{(\beta_i+\beta_j)^2}\biggl )  \nonumber \\ 
&+  \frac{1}{(\beta_i+\varepsilon_b)^2}\biggl ( \frac{R'''(-\beta_i)}{R'(\beta_i)} + \sum_j\frac{1}{2(\beta_i+\beta_j)^2} +  \sum_{j \neq i} \frac{1}{(\beta_i+\beta_j)^2}  \biggl )-\frac{2R''(-\beta_i)}{R'(-\beta_i)(\varepsilon_b+\beta_i)^3}  \nonumber\\
& - \frac{1}{\beta_i(\varepsilon_b-\beta_i)^3} + \frac{1}{\beta_i^2(\varepsilon-\beta_i)^2} - \frac{R''(-\beta_i)}{2R'(-\beta_i)\beta_i} \omega_{0,2}(\varepsilon_b,\beta_i)\bigg].
\end{align}
\end{theorem}
\textit{Proof.} By definition of our loop insertion operator, we simply have to carry out the $e_b$-derivative to finally obtain $\hat T_b \mathcal{F}^{(1)}=  \Omega_{1,1}(\varepsilon_b)$.  \\ 
Our strategy will be to carry out the action of $\hat T_b$ on $R'(0)$ and $R'(-\beta_i)$ separately. The result will be the following astonishing entanglement of both terms:
 \begin{align*}
&-\hat T_b \frac{\ln   ( R'(0)  )}{24}  = \frac{2}{3} \Omega^{BTR}_{1,1}(\varepsilon_b) \qquad \qquad \mathrm{\textbf{(A)}} \\ 
&\hat T_b  \frac{\mathfrak{R}_{\neq}}{24}-\hat T_b \frac{1}{24} \ln \biggl ( \prod_{i=1}^{2d} R'(-\beta_i) \biggl ) =   \Omega^{TR}_{1,1}(\varepsilon_b)  + \frac{1}{3} \Omega^{BTR}_{1,1}(\varepsilon_b)  \qquad \mathrm{\textbf{(B)}+\textbf{(C)}}
\end{align*}
implying that the $e_b$-derivative knows something about BTR, although it only acts on the usual $\mathcal{F}^{(1)}$ of topological recursion in the sense of Eynard and Orantin \cite{Eynard:2007kz}! \\ \\ 
The chain rule for the ordinary derivative $\partial_{e_b}$ acting on $R'(z)$ changes in a way that the inner function (the argument $z$) is naturally only affected, if the concrete value for $z$ has an explicit dependence  (e.g. $z=\pm \beta_i$, $z=\pm \varepsilon_k$). We write this dependence formally as $z(e_b)$ - for e.g. $z=0$ we have $z \neq z(e_b)$.
\begin{align}
\label{chainr}
\frac{\partial^2}{\partial z \partial{e_b}} R(z(e_b)) =\frac{\partial^2 }{\partial z^2}R(z(e_b)) \biggl (\frac{\partial z(e_b)}{\partial e_b} \biggl )+\biggl (\frac{\partial^2}{\partial z\partial e_b}R\biggl ) (z(e_b)).
\end{align}
\textbf{Step (A)}. For the derivative of $R'(0)$, the first summand in the chain rule vanishes and we obtain:
  \begin{align*}
&\frac{N}{r_b} \frac{\partial }{\partial e_b} R'(0) =\frac{N}{r_b} \biggl [ \frac{\partial }{\partial e_b}\frac{\partial}{\partial z}R(z)\biggl ]_{z=0}\\
&=\frac{N}{r_b} \frac{\partial }{\partial z }\bigg(\frac{\lambda}{N}\sum_k\bigg(\frac{r_k}{R'(\varepsilon_k)(\varepsilon_k+z)^2}\frac{\partial\varepsilon_k}{\partial e_b}+\frac{r_k}{R'(\varepsilon_k)^2(\varepsilon_k+z)}\frac{\partial R'(\varepsilon_k)}{\partial e_b}\bigg)\bigg)_{z=0}
 \end{align*}
Only the outer function $R(\varepsilon_k(e_b), R'(\varepsilon_k)(e_b))$ was affected by the ordinary derivative. To evaluate the terms in the parenthesis, the results from the interpolation formula in the proof of Lemma \ref{lem:2} can be inserted in complete analogy, giving
  \begin{align*}
&\frac{N}{r_b} \frac{\partial }{\partial z }\bigg[\frac{\lambda}{N}\frac{r_b}{R'(\varepsilon_b)}R'(z)\bigg(\frac{1}{R'(z)R'(-z)(z+\varepsilon_b)^2}\\
&+\sum_{i=1}^{2d}\frac{1}{(\beta_i-\varepsilon_b)^2}\frac{1}{R'(-\beta_i)R''(\beta_i)}\bigg(\frac{1}{(z-\beta_i)}+\frac{1}{(z+\beta_i)}\bigg)\bigg)\bigg]_{z=0}\\
=&16 \biggl ( \frac{\lambda}{R'(\varepsilon_b)}  (-\frac{\lambda}{8 (R'(0)) \varepsilon_b^3}
+\frac{\lambda R''(0)}{16(R'(0))^2\varepsilon_b^2}
-\sum_{i=1}^{2d} 
\frac{\lambda R'(0)}{8 \beta_i^2 R''(\beta_i) R'(-\beta_i)  (\varepsilon_b-\beta_i)^2} \biggl )
 \end{align*}
  Deriving instead the $\ln(R'(0))$ creates a prefactor $\frac{1}{R'(0)}$ and dividing by 24 gives the result as claimed. \\ \\ 
\textbf{Step (B)}.  We now have to admit an explicit dependence of the variable $z$ of the energy $e_b$ as it is the case for $z(e_b)=-\beta_i(e_b)$.  It arises from the general chain rule (\ref{chainr})
together with $ \partial_{e_b} (R'(\beta_i))=0$ (by definition of $\beta_i$) that
\begin{align*}
 &\frac{\partial^2 }{\partial z^2}R(z(e_b)) \biggl (\frac{\partial z(e_b)}{\partial e_b} \biggl )= -\biggl (\frac{\partial^2}{\partial z\partial e_b}R\biggl ) (z(e_b))
 \\ 
& \Rightarrow \quad R''(\beta_i)\frac{\partial \beta_i}{\partial e_b}=\frac{\lambda}{N}\sum_k\frac{2r_k}{R'(\varepsilon_k)(\varepsilon_k+\beta_i)^3}\frac{\partial \varepsilon_k}{\partial e_b}\\
&+\frac{\lambda}{N}\sum_k\frac{r_k}{R'(\varepsilon_k)^2(\varepsilon_k+\beta_i)^2}\frac{\partial R'(\varepsilon_k)}{\partial e_b} .
\end{align*}
The right hand side can be read off to be $\partial_{e_b} R'(-z)_{z=-\beta_i}$. We thus got a convergent expression of $\partial_{e_b}\beta_i$, whereas Lemma \ref{lem:2} fails: The needed derivative can be written in form of an involution identity - we have proved the following auxiliary relation:
\begin{lemma}
\label{inv2}
\begin{align}
&\partial_{e_b}R'(-\beta_i)= \partial_{e_b} \biggl [ R'(z)+ \frac{R''(-\beta_i)}{R''(\beta_i)} R'(-z) \biggl ]_{z=-\beta_i} 
\end{align} 
\end{lemma}
Again, the importance of the global involution $z \to -z$ is underlined in this relation.  The first summand on the rhs of Lemma \ref{inv2} will be of interest for step (B). We follow the calculation in step (A) and obtain a seemingly divergent expression, having however a convergent limit:
\begin{align*}
&\frac{N}{r_b} \lim_{z \to -\beta_i }\partial_{e_b} R'(z)\\
& =\lim_{z \to -\beta_i } R'(z) \cdot  \frac{\lambda}{R'(\varepsilon_b)} \frac{\partial }{\partial z } \bigg(\frac{1}{R'(z)R'(-z)(z+\varepsilon_b)^2}\\
&+\sum_{j=1}^{2d}\frac{1}{(\beta_j-\varepsilon_b)^2}\frac{1}{R'(-\beta_j)R''(\beta_i)}\bigg(\frac{1}{(z-\beta_j)}+\frac{1}{(z+\beta_j)}\bigg)\bigg) \\
&+ \lim_{z \to -\beta_i } R''(z) \cdot   \frac{\lambda}{R'(\varepsilon_b)}  \bigg(\frac{1}{R'(z)R'(-z)(z+\varepsilon_b)^2}\\
&+\sum_{j=1}^{2d}\frac{1}{(\beta_j-\varepsilon_b)^2}\frac{1}{R'(-\beta_j)R''(\beta_j)}\bigg(\frac{1}{(z-\beta_j)}+\frac{1}{(z+\beta_j)}\bigg)\bigg) 
\end{align*} 
 The Leibniz rule produces a term $\propto R''(z)$. We only have to perform the limit for the term $\propto R'(z)$, because the other one will be cancelled completely by the same term in step (C), eq. \ref{same} with an opposite sign. A series expansion yields the following preliminary result :
\begin{align}
\label{expansion}
&\lim_{z \to -\beta_i } R'(z) \cdot  \frac{\lambda}{R'(\varepsilon_b)} \frac{\partial }{\partial z } \bigg(\frac{1}{R'(z)R'(-z)(z+\varepsilon_b)^2} \nonumber \\
&+\sum_{j=1}^{2d}\frac{1}{(\beta_j-\varepsilon_b)^2}\frac{1}{R'(-\beta_j)R''(\beta_j)}\bigg(\frac{1}{(z-\beta_j)}+\frac{1}{(z+\beta_j)}\bigg)\bigg)\nonumber \\
&=-\frac{3}{R'(-\beta_i)R''(\beta_i) (\varepsilon_b-\beta_i)^4}-2\frac{R''(-\beta_i)}{R''(\beta_i)R'(-\beta_i)^2(\varepsilon_b-\beta_i)^3}+\frac{R'''(\beta_i)}{R''(\beta_i)^2R'(-\beta_i)(\varepsilon_b-\beta_i)^3}\nonumber \\
&+\frac{1}{(\varepsilon_b-\beta_i)^2}\bigg(\frac{R'''(\beta_i) R''(-\beta_i)}{2R''(\beta_i)^2R'(-\beta_i)^2}-\frac{R'''(\beta_i)^2 }{4R''(\beta_i)^3R'(-\beta_i)}+\frac{R''''(\beta_i)}{6R''(\beta_i)^2 R'(-\beta_i)}\nonumber \\
&\quad +\frac{R'''(-\beta_i)}{2R''(\beta_i)R'(-\beta_i)^2}-\frac{R''(-\beta_i)^2}{R''(\beta_i)R'(-\beta_i)^3}\bigg)-\sum_j\frac{1}{(\beta_j-\varepsilon_b)^2}\frac{1}{R'(-\beta_j)R''(\beta_j)}\frac{1}{(\beta_i+\beta_j)^2} \nonumber \\
&-\sum_{i\neq j}\frac{1}{(\beta_j-\varepsilon_b)^2}\frac{1}{R'(-\beta_j)R''(\beta_j)}\frac{1}{(\beta_i-\beta_j)^2}.
\end{align}
This is not far from $\omega_{1,1}^{TR}$ compared with Proposition \ref{om11} - we observe just two wrong prefactors and a remainder that has to be treated separately. In order to generate the known form of $\omega_{1,1}^{TR}$, consider the following expansion of $\frac{R''(z)}{R'(z)}$ around its single poles $\beta_i$ and $-\varepsilon_k$.
\begin{align} \label{pfe}
\frac{R''(z)}{R'(z)}&=\sum_i\frac{1}{z-\beta_i}-2\sum_k\frac{1}{z+\varepsilon_k}\\\nonumber
\rightarrow\quad \frac{R'''(z)}{R'(z)}-\frac{R''(z)^2}{R'(z)^2}&=-\sum_i\frac{1}{(z-\beta_i)^2}+2\sum_k\frac{1}{(z+\varepsilon_k)^2}
\end{align}
We perform another expansion around $z=\beta_i$ of 
\begin{align*}
\frac{R'''(z)}{R'(z)}-\frac{R''(z)^2}{R'(z)^2}=-\frac{1}{(z-\beta_i)^2}+\frac{R''''(\beta_i)}{3R''(\beta_i)}-\frac{R'''(\beta_i)^2}{4R''(\beta_i)^2}
\end{align*}
giving the identity of desire
\begin{align}
\label{id2}
\frac{R''''(\beta_i)}{3R''(\beta_i)}-\frac{R'''(\beta_i)^2}{4R''(\beta_i)^2}=-\sum_{i\neq j}\frac{1}{(\beta_i-\beta_i)^2}+2\sum_k\frac{1}{(\beta_i+\varepsilon_k)^2}
\end{align}
Inserting (\ref{id2}) into eq. (\ref{expansion}) and deriving $\ln(\prod_i R'(-\beta_i))$ instead gives a sum over all ramification points and a prefactor $\frac{1}{R'(-\beta_i)}$, so that we simply end up with
\begin{align*}
& 24 \cdot \omega_{1,1}^{TR}+ \mathcal{R} \\
& \mathcal{R}=- \sum_{i=1}^{2d}\frac{1}{(\beta_i-\varepsilon_b)^2}\frac{1}{R'(-\beta_i)R''(\beta_i)}\bigg(\frac{R''(-\beta_i)^2}{R'(-\beta_i)^2}+\sum_j\frac{1}{(\beta_i+\beta_j)^2}\\
&+2\sum_k\frac{1}{(\varepsilon_k+\beta_i)^2}+2\frac{R''(-\beta_i)}{R'(-\beta_i)(\varepsilon_b-\beta_i)}\bigg).
\end{align*}
that has to be further simplified. This will be possible together with the results of the second summand of the rhs of Lemma \ref{inv2} that has will be carefully evaluated in step (C).  \\ \\ 
\textbf{Step (C)}. To do that, we continue with $\frac{N}{r_b} \partial_{e_b}\frac{R''(-\beta_i)}{R''(\beta_i)} R'(-z)$ evaluated at $z=-\beta_i$. The Leibniz rule prevents us from an evaluation of $\partial_{e_b}\frac{R''(-\beta_i)}{R''(\beta_i)}$ because we have a zero as prefactor coming from $ R'(-z)_{z=-\beta_i}$. With our knowledge up to know, we cannot directly evaluate $\frac{N}{r_b} \partial_{e_b} R'(-z)$   at $z=-\beta_i$. This can be seen if one looks at the limit process performed in step (B), eq. (\ref{expansion}), but for $\lim_{z \to \beta_i}$ - a divergent expression occurs (for the same reason we already had to introduce Lemma \ref{inv2}). For a change of strategy, we derive Lemma \ref{lem:1} with respect to $z$ and insert $z=-\beta_i$ giving a new relation:
 \begin{align*}
 &R''(-\beta_i) \partial_{e_b}R(-z)_{z=-\beta_i}-R'(-\beta_i) \partial_{e_b}R'(-z)_{z=-\beta_i}-R''(\beta_i) \partial_{e_b}R(z)_{z=-\beta_i}\\
&= -\frac{2r_b}{R'(\varepsilon_b)N}\biggl ( \frac{1}{(\varepsilon-\beta_i)^3}- \frac{1}{(\varepsilon+\beta_i)^3} \biggl )
\end{align*}  
or in our desired form 
 \begin{align*}
 &\frac{N}{r_b} \biggl [ \frac{R''(-\beta_i)^2}{R''(\beta_i)R'(-\beta_i)} \partial_{e_b}R(-z)_{z=-\beta_i}-\frac{R''(-\beta_i)}{R'(-\beta_i)} \partial_{e_b}R(z)_{z=-\beta_i}\\
&+\frac{2R''(-\beta_i)}{R''(\beta_i)R'(-\beta_i)}\frac{r_b}{R'(\varepsilon_b)N}\biggl ( \frac{1}{(\varepsilon-\beta_i)^3}- \frac{1}{(\varepsilon+\beta_i)^3} \biggl ) \biggl ]= \frac{N}{r_b} \frac{R''(-\beta_i)}{R''(\beta_i)} \partial_{e_b}R'(-z)_{z=-\beta_i}
\end{align*}  
 Let us evaluate this expression term by term. Lemma \ref{lem:1} states that the first term $\frac{N}{r_b} R''(-\beta_i) \partial_{e_b}R(-z)_{z=-\beta_i}$ reads
\begin{align*}
 \frac{1}{R'(\varepsilon_b)} \frac{R''(-\beta_i)^2}{R'(-\beta_i)^2R''(\beta_i)}\biggl ( \frac{1}{(\varepsilon-\beta_i)^2}+\frac{1}{(\varepsilon+\beta_i)^2} \biggl )
\end{align*}
since $\frac{R'(-z)}{R'(z)} \partial_{e_b} R(z)$ vanishes at $z=-\beta_i$. The second summand can be obtained exactly like in step (B) for  $\lim_{z \to- \beta_i }\partial_{e_b} R'(z)$ with 
\begin{align}
\label{same}
\frac{R''(-\beta_i)}{R'(-\beta_i)} &\lim_{z \to- \beta_i }  \frac{\lambda}{R'(\varepsilon_b)}R'(z)\bigg[\frac{1}{R'(z)R'(-z)(z+\varepsilon_b)^2}\\
&+\sum_{i=1}^{2d}\frac{1}{(\beta_i-\varepsilon_b)^2}\frac{1}{R'(-\beta_i)R''(\beta_i)}\bigg(\frac{1}{(z-\beta_i)}+\frac{1}{(z+\beta_i)}\bigg)\bigg]
\end{align} 
As announced, this term perfectly cancels the omitted term  $\propto R''(z)$ in (B) due to its different prefactors.  Therefore, all the remaining terms of the proof read:
\begin{align*}
&\mathcal{R} + \frac{1}{R'(\varepsilon_b)} \sum_{i=1}^{2d}\biggl [  \frac{R''(-\beta_i)^2}{R'(-\beta_i)^2R''(\beta_i)}\biggl ( \frac{1}{(\varepsilon-\beta_i)^2}+\frac{1}{(\varepsilon+\beta_i)^2} \biggl ) \\
&-\frac{2R''(-\beta_i)}{R''(\beta_i)R'(-\beta_i)} \biggl ( \frac{1}{(\varepsilon-\beta_i)^3}- \frac{1}{(\varepsilon+\beta_i)^3} \biggl )  \biggl ] \\
&= \frac{1}{R'(\varepsilon_b)} \sum_{i=1}^{2d} \frac{ 1}{R'(-\beta_i)R''(\beta_i)}\bigg( - \frac{1}{(\beta_i-\varepsilon_b)^2} \biggl ( \sum_j\frac{1}{(\beta_i+\beta_j)^2}+2\sum_k\frac{1}{(\varepsilon_k+\beta_i)^2}\biggl ) \\
&+  \frac{1}{(\beta_i+\varepsilon_b)^2}\biggl ( \frac{R'''(-\beta_i)}{R'(\beta_i)} + \sum_j\frac{1}{(\beta_i+\beta_j)^2}-2\sum_k\frac{1}{(\varepsilon_k-\beta_i)^2} \biggl )-\frac{2R''(-\beta_i)}{R'(-\beta_i)(\varepsilon_b+\beta_i)^3}\bigg)
\end{align*}
where we used again the partial fraction decomposition (\ref{pfe}), now evaluated at $z=-\beta_i$, giving
\begin{align*}
\frac{R''(-\beta_j)^2}{R'(-\beta_j)^2} = \frac{R'''(-\beta_j)}{R'(-\beta_j)}+\sum_i\frac{1}{(\beta_j+\beta_i)^2}-2\sum_k\frac{1}{(\varepsilon_k-\beta_j)^2}
\end{align*}
 Our final aim is to prove that this remainder behaves like $ \frac{1}{3} \Omega^{BTR}_{1,1}(\varepsilon_b)$. Some terms directly cancel due to
\begin{align*} 
0=\sum_i\frac{1}{R'(-\beta_i)R''(\beta_i)}\bigg(\frac{1}{(\varepsilon_a-\beta_i)^2(\beta_i+\varepsilon_b)^2}-\frac{1}{(\varepsilon_a+\beta_i)^2(\varepsilon_b-\beta_i)^2}\bigg)
\end{align*}
This relation can be obtained with the expansion (\ref{pfe0}) evaluated at $(z,w)=(\varepsilon_a,\varepsilon_b)$.
However, further cancellations differ dependent on the choice $d=1$ or $d>1$. It is reasonable to split the proof at this point. \\ \\ 
\textbf{(C1)}  For $d=1$, we find two equivalent expression of a constituent of $\Omega^{BTR}_{1,1}(\varepsilon_b)$, namely
\begin{align*}
&\sum_{i=1}^2 \frac{1}{R'(-\beta_i)R''(\beta_i)} \bigg(  \frac{1}{(\beta_i+\varepsilon_b)^2}- \frac{1}{(\beta_i-\varepsilon_b)^2} \biggl ) \sum_j\frac{1}{(\beta_i+\beta_j)^2} \\
& = \sum_{i=1}^2 \frac{1}{R'(-\beta_i)R''(\beta_i)}\frac{2R''(-\beta_i)}{R'(-\beta_i)(\varepsilon_b+\beta_i)^3} = \frac{1}{2R'(0)^2\varepsilon_b^3}
\end{align*}  
Moreover, we find for general $d$:
\begin{align*}
	&\frac{R''(-z)}{R'(-z)^2R'(z)(z+w)}=\\
&\frac{R''(w)}{R'(w)^2R'(-w)(z+w)}+\sum_i\frac{1}{R''(\beta_i)R'(-\beta_i)}\bigg(\frac{1}{(z+\beta_i)^2(w-\beta_i)}\\
&	-\frac{1}{(z+\beta_i)(w-\beta_i)^2}-\frac{R''(-\beta_i)}{R'(-\beta_i)(z+\beta_i)(w-\beta_i)}+\frac{R''(-\beta_i)}{R'(-\beta_i)(z-\beta_i)(w+\beta_i)}\bigg)
\end{align*}
Deriving wrt $w$ and setting $(z,w)=(0,\varepsilon_b)$ gives the last missing piece in  $\frac{R''(0)}{48R'(0)^3\varepsilon_b^2}$.
Thus, we finally reduced our last part to something that can be easily shown to be equal to $  \frac{1}{3} \Omega^{BTR}_{1,1}(\varepsilon_b)$ (in the combinatorial limit, a check with computer algebra could be performed) and the proof is finished.  $\qed_1$ \\ \\ 
\textbf{(C2)} For $d \neq 1$, some relations change.  Derive 
\begin{align}\label{pfe2}
\frac{1}{R'(z)R'(-z)(z+\varepsilon_b)^2}=\sum_i\frac{1}{R'(-\beta_i)R''(\beta_i)}\bigg(\frac{1}{(z-\beta_i)(\beta_i+\varepsilon_b)^2}-\frac{1}{(z+\beta_i)(\varepsilon_b-\beta_i)^2}\bigg)
\end{align}
(see proof of Lemma \ref{lem:2}) with respect to $z$ and evaluate the result at $z=0$. This proves for general $d$ that
\begin{align*}
&\sum_i \frac{1}{R'(-\beta_i)R''(\beta_i)\beta_i^2} \bigg(  \frac{1}{(\beta_i+\varepsilon_b)^2}- \frac{1}{(\beta_i-\varepsilon_b)^2} \biggl ) = \frac{2}{R'(0)^2\varepsilon_b^3}
\end{align*}
Setting $d\neq1$, the simplification in step (C1) thus only holds for $i=j$ in $\sum_{i,j} \frac{1}{(\beta_i+\beta_j)^2}$, giving already the hint that $\mathfrak{R}_{\neq}$ should contain a double-sum without diagonal. Moreover note that
\begin{align*}
& \frac{1}{2R'(0)^2\varepsilon_b^3} \neq \sum_{i=1}^{2d} \frac{1}{R'(-\beta_i)R''(\beta_i)}\frac{2R''(-\beta_i)}{R'(-\beta_i)(\varepsilon_b+\beta_i)^3}  
\end{align*}  
for general $d$. Together with the relations in step (C1), we extract $ \frac{1}{3} \Omega^{BTR}_{1,1}(\varepsilon_b)$ from the rest term and find something non-vanishing which we interpret as the derivative of the compensation term announced in the theorem, eq. (\ref{simpl})
It is not possible with the known techniques to guess the primitive of $\frac{\mathfrak{R}_{\neq}}{24}$ with respect to $\hat T_b$. We only deliver the first order in a perturbative series giving already a glimpse of this rest term:
\begin{align*}
\mathfrak{R}_{\neq} =-\frac{ \lambda}{N} \sum_{k,l, k\neq l}^d \frac{r_kr_l}{(e_k+e_l)^2} + \mathcal{O}(\lambda^2)
\end{align*}
Apart from this compensation term, we were able to construct the complete $\omega_{1,1}(\varepsilon_b)$ using the loop insertion operator. $\qed_2$

\section{Conclusion and Outlook}
In this paper, we have shown in which way the genus-one free energy $\mathcal{F}^{(1)}$ differs from its generic form in usual topological recursion, if additional holomorphic contributions occur: The concrete blob for $\omega_{1,1}$ in the quartic Kontsevich model, having poles at the fixed point $0$ of the global symmetry $x(z)=-y(-z)$ of the spectral curve behind, influences the expression for $\mathcal{F}^{(1)}$. We interpreted this object as a primitive with respect to a boundary creation/loop insertion operator in blobbed topological recursion. This operator became the main tool in the lengthy, but mostly elementary proof. 

The idea of what to be proved came into being when we compared the combinatorial limit of the quartic Kontsevich model (the external field becomes a scalar) with the results of the Hermitian one-matrix model for quadrangulations. Using the perturbative approach via expansion into ribbon graphs with tetravalent vertices, we recognized the role of $\ln(R'(0))$ for $\mathcal{F}^{(1)}$. Moreover, it turned out that the Bergman $\tau$-function, non-vanishing in the one-matrix model and many more, can be neglected in our blobbed topological recursion, although the solution of the determining ODE is non-zero. Adding this term artificially to $\mathcal{F}^{(1)}$, we obtained a primitive of the pure TR-part of $\omega_{1,1}$. Its power series expansion revealed that the pure TR constituents, when we set any holomorphic add-on's to zero, are generating functions of only the \textit{bipartite} quadrangulations in the sense of \cite{Carrell:2014}, which holds also for higher $g$. Therefore, the investigation of the free energy accidentally gave us an interpretation of our blobbed TR in terms of concepts in enumerative geometry!  

However, we were faced with an unknown compensation term $\mathfrak{R}_{\neq}$ that vanishes in the combinatorial limit, but not for an arbitrary number of ramification points. 
A perturbative approach to this term failed - it seems nearly impossible to guess the candidate after having excluded the obvious ones. 
Furthermore, something comparable was not found in the literature. \\ \\ 
 \textit{We therefore invite the reader with some expertise in topological recursion to make a suggestion, where the compensation term might have its origin. An approximative solution is already given in the main result. We are grateful for any hint that may explain this final step of the proof!}

\appendix
\section{Perturbative Expansion of $\mathcal{F}^{(1)}$}
\label{appA}
At the beginning of our calculations, we tried to guess the exact compensation term $\mathfrak{R}_{\neq}$ by expanding $\mathcal{F}^{(1)}$ into closed ribbon graphs up to order $\lambda^2$ as introduced in Chapter \ref{ch:setup}. We take a look at the elements of $\mathfrak{G}^{1,v}_\emptyset$ for general $d$ to get a feeling  for the compensation term. \\ \\ 
$\mathcal{O}(\lambda)$: For $v=1$, only one graph contributes with 
\begin{align*}
\Gamma=\frac{(-\lambda)}{4N^2}\sum_{k}^d\frac{r_k}{(2e_k)^2}
\end{align*}
Hence, we see that only the expansion of $R'(0)$ term in $\mathcal{F}^{(1)}$ gives a the expected contribution (one face to be integrated out):
 \begin{align*}
& \frac{(-\lambda)}{12N}  \textcolor{red}{\sum_{k,l=1}^d\frac{r_kr_l}{(e_k+e_l)^2}}+  \frac{(-\lambda)}{24N}   \sum_{k=1}^d\frac{r_k}{e_k^2} = \frac{(-\lambda)}{12N}  \textcolor{red}{\sum_{k,l=1, k \neq l}^d\frac{r_kr_l}{(e_k+e_l)^2}}+  \frac{(-\lambda)}{16N}   \sum_{k=1}^d\frac{r_k}{e_k^2} 
\end{align*}
Setting $d=1$ gives $\frac{-\lambda}{4} \frac{1}{4e^2}$ (the weights $(2e)^{2k}$ were omitted in the above generating series), because the sum $k \neq l$ vanishes. An obvious candidate for $\mathfrak{R}_{\neq}$ is
\begin{align*}
\sum_{k,l, k \neq l} \frac{\lambda r_kr_l}{(\varepsilon_k+\varepsilon_l)^2}
\end{align*}
We know
\begin{align*}
\frac{\partial \varepsilon_a}{\partial e_b}=&\frac{\delta_{a,b}}{ R'(\varepsilon_a)}-\frac{\lambda}{N}\frac{r_b}{R'(\varepsilon_b)}\sum_{i=1}^{2d}\bigg(\frac{1}{(\beta_i+\varepsilon_b)^2}+\frac{1}{(\beta_i-\varepsilon_b)^2}\bigg)\frac{1}{(\varepsilon_a-\beta_i)R'(-\beta_i)R''(\beta_i)}
\end{align*}
and assume a chain rule for the insertion operator:
\begin{align*}
\frac{\partial }{\partial e_b}\sum_{k,l, k \neq l} \frac{r_kr_l}{(\varepsilon_k+\varepsilon_l)^2} = - \sum_{k,l, k \neq l}  \frac{\partial \varepsilon_k}{\partial e_b} \frac{2r_kr_l}{(\varepsilon_k+\varepsilon_l)^3}  - \sum_{k,l, k \neq l}  \frac{\partial \varepsilon_l}{\partial e_b} \frac{2r_kr_l}{(\varepsilon_k+\varepsilon_l)^3}
\end{align*}
Both terms are of course equal due to the symmetry. However, this is not the remainder of what has to be compensated. It does not match with the surplus structure arising from the derivative of $R'(-\beta_i)$.  \\ \\ 
$\mathcal{O}(\lambda^2)$: For $v=2$, we have the contributions
\begin{align*}
&\frac{\ln(R'(0))}{24}[\lambda^2]=-\frac{1}{48N^2} \sum_{k,l=1}^N r_kr_l\biggl ( \frac{2}{(e_k+e_l)^2e_l^2}+\frac{1}{e_k^2e_l^2}\\
&+\frac{4}{(e_k+e_l)e_l^3}\biggl ):=c_{0,1}+c_{0,2}+c_{0,3}
\end{align*}
where $d=1$ gives $\frac{28}{24}$ and
 \begin{align*}
&\frac{\ln(\Pi_iR'(-\beta_i))}{24}[\lambda^2]=-\frac{1}{4N^2}\sum_{k,l=1}^N\frac{r_kr_l}{(e_k+e_l)^4} - \sum_{k,l,n=1}^N  r_kr_l \biggl (-\frac{1}{3N^2} \frac{1}{(e_k+e_l)^3(e_n+e_l)}\\
&   -\frac{1}{8N^2} \frac{1}{(e_n+e_l)^2(e_n+e_k)^2}  \biggl ):= c_{\beta,1}+ c_{\beta,2}+ c_{\beta,3}
\end{align*}
where $d=1$ gives $\frac{17}{24}$, thus we have the expected rational number $\frac{15}{8}$ we encountered in \ref{f1numbers}. \\ \\ 
The following graphs give a contribution:
\begin{align*}
&\Gamma_1=-\frac{\lambda^2}{N^2} \sum_{k,l} \frac{r_kr_l}{8e_k^3(e_k+e_l)} \qquad \quad
&\Gamma_2=-\frac{\lambda^2}{4N^2} \sum_{k,l} \frac{r_kr_l}{4e_ke_l(e_k+e_l)^2}\\
&\Gamma_3=-\frac{\lambda^2}{2N^2} \sum_{k,l} \frac{r_kr_l}{4e_k^2(e_k+e_l)^2} \qquad 
&\Gamma_4=-\frac{\lambda^2}{8N^2} \sum_{k,l} \frac{r_kr_l}{(e_k+e_l)^4}
\end{align*}
 It is simple to verify the relation
\begin{align*}
 c_{0,1}+c_{0,2}= -\frac{\lambda^2}{N^2}\sum_{k,l} \biggl ( \frac{r_kr_l}{12e_l^2(e_k+e_l)^2} +\frac{r_kr_l}{24(e_k+e_l)^3e_l} \biggl )
\end{align*}
Therefore, we again observe the behaviour $\frac{3}{2}(c_{0,1}+c_{0,2}+c_{0,3})=\Gamma_1+\Gamma_2+\Gamma_3$ (only $\Gamma_4$ is left untouched - the bipartite one!). It is the perturbative confirmation of step (A) in the main proof that $\frac{\ln(R'(0))}{24}$ is responsible for $\frac{2}{3} \Omega^{BTR}_{1,1}$, meaning all the non-bipartite Feynman graphs. One directly sees that Conjecture \ref{bipconj} holds also for $d>1$ regarding the generation of bipartite decorated maps/ribbon graphs. To compensate the $\frac{3}{2}$, use again the trick
\begin{align*}
 c_{0,3}+c_{\beta,2}|_{k=l}=\Gamma_1 \qquad c_{\beta,3}|_{k=l}=\Gamma_4\qquad
\frac{3}{2}(c_{0,1}+c_{0,2})=\Gamma_2+\Gamma_3
\end{align*}
we thus conclude that the rest term to $\mathcal{F}^{(1)}$ reads at order $\lambda^2$: $-\frac{1}{2}(c_{0,1}+c_{0,2})+(c_{\beta,2}+c_{\beta,3})_{k=l}$. At this point, there are plenty representations of this expression. It becomes nearly impossible to guess the complete compensation term from perturbation theory. With the numerical support of a computer program of Jakob Lindner, we could exclude some promising candidates for the compensation term also at order $\mathcal{O}(\lambda^2)$.

\section{On the Bergman projective connection}
\label{appB}
The role of the reflected standard bidifferential in 
  $\omega_{0,2}(u,z)=\frac{du\,dz}{(u-z)^2}+\frac{du\,dz}{(u+z)^2}$
in the kind of blobbed topological recursion is still not completely clarified. We mention the appearance of the same $\omega_{0,2}(u,z)$ in the context of spin Hurwitz numbers \cite{Giacchetto:2021} based on another kind of reflection symmetry in the spectral curve. The main difference is, however, that we only include the standard Bergman kernel without holomorphic additions into the recursion kernel and somehow treat the reflected Bergman kernel rather as a first blob $\phi_{0,2}(u,z)=B(u,-z)$. This is not common in the general framework of BTR in \cite{Borot:2015hna}. In the context of enlarged $\omega_{0,2}$ and $\mathcal{F}^{(1)}$, the concept of the Bergman projective connection $S_B$ shows up. It is introduced via a local expansion nearby $q \to p$
 \begin{align*}
	\frac{\omega_{0,2}(p,q)}{d\zeta(p)d\zeta(q)} =\frac{1}{(\zeta(p)-\zeta(q))^2}+\frac{S_B(p)}{6} + \mathcal{O}((\zeta(p)-\zeta(q))^1) \; .
\end{align*}
in a proper local variable. On the one hand, it shows up in the context of symplectic invariances. Let $\hat{\omega}^{TR}_{1,1}(z)$ be the meromorphic differential created by topological recursion, but with exchanging the role of $x$ and $y$ in the underlying spectral curve. Then the following relation holds \cite{Eynard:2007kz}:
\begin{theorem}
\label{PropEO}
 \begin{align*}
	&\omega_{1,1}^{TR}(z)+\hat{\omega}_{1,1}^{TR}(z) =  \frac{\partial}{\partial z} \frac{1}{24x'y'} \biggl (2S_B(z)+\frac{x''y''}{2x'y'}-\frac{x'''}{x'}+\frac{x''^2}{x'^2}+ x \leftrightarrow y \biggl )
\end{align*}
\end{theorem}
An easy calculation shows the following surprising result for our result in blobbed topological recursion, completely coinciding with the formula from above:
\begin{proposition}
\begin{align*}
	&R'(z)\Omega_{1,1}(z)-R'(-z)\Omega_{1,1}(-z)= \frac{\partial}{\partial z}\frac{\lambda}{24R'(z)R'(-z)} \\
& \times  \bigg[\frac{3}{z^2}-\frac{R'''(z)}{  R'(z)}-\frac{R'''(-z)}{  R'(-z)}+\frac{R''(z)^2}{  R'(z)^2}+\frac{R''(-z)^2}{ R'(-z)^2}-\frac{R''(z)R''(-z)}{  R'(z)R'(-z)}\bigg].
\end{align*}
\end{proposition}
 We can read off $S_B(z)=\frac{3}{2z^2}$ in full compliance with the behaviour of $\lim_{u \to z} \phi_{0,2}(u,z) = \frac{S_B(z)}{6}$, in particular we can decompose into
\begin{align}
\label{SB}
&R'(z)\Omega^{BTR}_{1,1}(z)-R'(-z)\Omega^{BTR}_{1,1}(-z)= \frac{\partial}{\partial z}\frac{\lambda S_B(z)}{12R'(z)R'(-z)}  
\end{align}
and the rest for the normalised solution $R'(z)\Omega^{TR}_{1,1}(z)$ always implying $S_B=0$ for only the standard bidifferential. First, we notice that taking the complete $\omega_{0,2}$ into the recursion kernel as in many other models does not produce a satisfying result, the expression $R'(z)\Omega^{TR}_{1,1}(z)-R'(-z)\Omega^{TR}_{1,1}(-z)$ would vanish what should not happen due to the generality of the result in TR. The remarkable conclusion of eq. (\ref{SB}) is that $\phi_{0,2}$ or rather the \textit{Bergman projective connection $S_B(z)$ already contains the information about the blob $\mathcal{H}_z \omega_{1,1}(z)$   in order to fulfil a pure TR result}, Thm. \ref{PropEO}.
 $S_B$ also shows up in its primitive $\mathcal{F}^{(1)}$ if one defines the Bergman $\tau$-function as in the original works \cite{Eynard:2002}
 \begin{align}
\label{taukk}
\partial_{b_i} \ln[\tau_B(b_1,...,b_{2d})]=\frac{S_B(\zeta_i)_{\zeta_i=0}}{6} \qquad x(\beta_i) =:b_i
\end{align}
with a local variable $\zeta_i(z)=\sqrt{x(z)-b_i}$ nearby the simple ramification points.
 
An interesting observation regards the genus-one free energy of the Hermitian two-matrix model \cite{Eynard:2002b}: The creation operator belonging to the 2MM is a derivative with respect to one of the potentials, $V_1$. Its application yields the usual TR result of $\omega_{1,1}$ and an additional term which is absorbed via the Bergman $\tau$-function understood as in eq. (\ref{taukk}). This term reads
\begin{align*}
\frac{\partial \tau_B}{\partial V_1 }  =(d_2+1)\sum_{i=1}^{d_2+1} \frac{1}{y'(\beta_i)x''(\beta_i)\beta_i^2(z-\beta_i)^2} \quad .
\end{align*}
Taking into account our additional $\phi_{0,2}$, the topological recursion formula produces a term of exactly the same structure in $ \Omega^{BTR}_{1,1}(\varepsilon_b)$.  

All these coincidences suggest to focus the future research also on the possibility that our model covered by BTR might be reformulated in a proper language that only requires elements of pure topological recursion. In this paper, we recognised that no Bergman $\tau$-function is necessary for $\mathcal{F}^{(1)}$, however we had to add $\ln(R'(0))$ in order to produce  $\mathcal{H}_z \omega_{1,1}(z)$. One implication of our Bergman projective connection $S_B$ might be that the blobs are only a surplus structure. It is under current investigation, if a suitable change of e.g. the genus of the spectral curve might turn the blobbed topological recursion of the QKM into a pure TR model (there are prime examples such as the $O(n)$ model that does not need blobs for a $g=1$ spectral cuve \cite{Borot:2009ia}).

\bibliographystyle{halpha-abbrv}
\bibliography{free_en}

\end{document}